\documentclass[twocolumn,epjc3]{svjour3}
\usepackage{eso-pic}
\usepackage[T1]{fontenc}
\usepackage[utf8]{inputenc}
\usepackage[dvipsnames]{xcolor}
\usepackage{microtype}
\usepackage{graphicx}
\usepackage{amsmath}
\usepackage{amssymb}
\usepackage{booktabs}
\usepackage{tabularx}
\usepackage{array}
\usepackage{arydshln}
\usepackage{placeins}
\usepackage{cite}
\usepackage{hyperref}
\usepackage{newtxtext,newtxmath,relsize}
\hypersetup{
  colorlinks=true,
  linkcolor=ForestGreen,
  citecolor=ForestGreen,
  filecolor=ForestGreen,
  urlcolor=ForestGreen,
  pdftitle={Quadruple-Higgs Boson Production at the LHC},
  pdfauthor={Andreas Papaefstathiou and Gilberto Tetlalmatzi-Xolocotzi},
  pdfsubject={Quadruple-Higgs boson production at the high-luminosity LHC},
  pdfkeywords={Higgs boson, self-couplings, high-luminosity LHC}
}

\journalname{Eur. Phys. J. C}
\graphicspath{{figures/}}
\newcolumntype{Y}{>{\raggedright\arraybackslash}X}

\newcommand{\preprintA}{SI-HEP-2026-18}

\newif\ifshowpreprints
\showpreprintstrue

\newcommand{\preprintfont}{\normalfont\rmfamily\footnotesize}
\newcommand{\preprintcolor}{black}
\newlength{\preprintTopMargin}
\newlength{\preprintRightMargin}
\newcommand{\PutPreprintFirstPage}{%
  \ifshowpreprints
  \AddToShipoutPictureFG*{%
    \put(
      \LenToUnit{\dimexpr\paperwidth-\preprintRightMargin\relax},
      \LenToUnit{\dimexpr\paperheight-\preprintTopMargin\relax}
    ){%
      \makebox[0pt][r]{%
        \preprintfont \color{\preprintcolor}%
        \shortstack[r]{\preprintA\\\preprintB\\\preprintC}%
      }%
    }%
  }%
  \fi
}

\begin{document}

\PutPreprintFirstPage

\title{Quadruple-Higgs Boson Production at the LHC}

\author{Andreas Papaefstathiou\thanksref{i1,e1} \and
        Gilberto Tetlalmatzi-Xolocotzi\thanksref{i2,e2}}

\institute{\hypertarget{i1}{}Department of Physics, Kennesaw State University, 830 Polytechnic Lane, Marietta, GA 30060, USA \label{i1} \and
           \hypertarget{i2}{}Theoretische Physik 1, Center for Particle Physics Siegen (CPPS), Universit\"at Siegen, Walter-Flex-Str. 3, 57068 Siegen, Germany \label{i2}}

\thankstext{e1}{\href{mailto:apapaefs@kennesaw.edu}{apapaefs@kennesaw.edu}}
\thankstext{e2}{\href{mailto:gtx@physik.uni-siegen.de}{gtx@physik.uni-siegen.de}}

\date{August 2026}
\maketitle

\begin{abstract}
We investigate the production of \textit{four} Higgs bosons via gluon fusion
at the high-luminosity LHC. We construct a phenomenological analysis of the
eight-$b$-jet final state and use a multivariate analysis to derive expected
simultaneous constraints on the triple and quartic Higgs boson self-coupling
modifications. At the 95\% confidence level, the resulting region projects onto
$-10.3<c_3<11.2$ and $-222<d_4<234$, where $c_3=\kappa_3-1$ and
$d_4=\kappa_4-1$. We also examine how quadruple-Higgs boson production can enter a
triple-Higgs boson signal region requiring at least six $b$-tagged jets.

In addition, we consider two extensions of the Standard Model scalar sector:
a one-real-singlet extension yielding the direct resonant process
\(pp\rightarrow S\rightarrow hhhh\), and a two-real-singlet extension yielding
the cascade \(pp\rightarrow h_3\rightarrow h_2h_2\rightarrow (hh)(hh)\), where
$S$, $h_2$ and $h_3$ are new scalar resonances. For these processes, we derive
expected upper limits on the quadruple-Higgs boson production cross sections as
functions of the new scalar masses.
\end{abstract}

\keywords{Higgs boson \and self-couplings \and high-luminosity LHC}

\section{Introduction}\label{sec:intro}

The scalar sector of the Standard Model (SM) remains poorly
constrained. The high-luminosity phase of the CERN Large Hadron Collider
(HL-LHC), which is set to commence in the 2030s, will provide several avenues
in the form of rare multi-Higgs boson final states to explore this sector. The
integrated luminosity expected by the end of its operation,
\(\mathcal{L}=3000~\mathrm{fb}^{-1}\) per experiment, is expected to
constrain the Higgs boson's trilinear coupling to the \(\mathcal{O}(50\%)\)
level, provided the SM value is realized in nature~\cite{ATLAS:2025wdq,Collaboration:2928096}.
In contrast, measuring the quartic self-coupling will remain extremely
challenging~\cite{Plehn:2005nk}, since its primary direct probe, triple-Higgs boson production via
gluon fusion~\cite{Abouabid:2024gms}, is extremely rare, with a cross section of
\(\sigma_{hhh}\simeq0.1~\mathrm{fb}\) at 14~TeV~\cite{Maltoni:2014eza,deFlorian:2019app},
and competes with comparatively large multijet QCD backgrounds. Furthermore,
the dependence of triple-Higgs boson production on the quartic coupling is
small compared with its dependence on the trilinear coupling; see, e.g.,
Ref.~\cite{Papaefstathiou:2019ofh}. Projected HL-LHC intervals from
\(gg\rightarrow hhh\), particularly in the six-\(b\)-jet final state, can
therefore extend to quartic-coupling modifications of $\mathcal{O}(100)$ when
interpreted in a simultaneous coupling plane~\cite{ATLAS:2025cae}.

To maximize the information on the quartic self-coupling available at the
HL-LHC, it is useful to explore a broad set of triple-Higgs boson final states
in addition to \(hhh\rightarrow6b\)~\cite{Papaefstathiou:2019ofh,Papaefstathiou:2023uum,Stylianou:2023tgg,ATLAS:2024xcs,ATLAS:2025cae,Fuks:2025gjv,CMS:2026xkc,Papaefstathiou:2020lyp,Karkout:2024ojx,Papaefstathiou:2025meh,Chiang:2025ecn,Procter:2026rlk}.
These include \(hhh\rightarrow4b+2\gamma\)~\cite{Papaefstathiou:2015paa,Fuks:2015hna,Chen:2015gva,CMS:2025jkb},
\(hhh\rightarrow4b+2\tau\)~\cite{Fuks:2015hna,Fuks:2017zkg},
\(hhh\rightarrow2b+4\tau\)~\cite{Stylianou:2023tgg,Dong:2025lkm}, and
\(hhh\rightarrow2b+4W\)~\cite{Kilian:2017nio}. Indirect constraints, e.g.\
through loop effects in \(gg\rightarrow h\) or \(gg\rightarrow hh\), can
provide additional information while introducing some model
dependence~\cite{Bizon:2018syu,Bizon:2024juq,Heinrich:2024dnz,Haisch:2025pql}.

Here, we investigate the prospect of employing the extremely rare gluon-fusion-initiated
quadruple-Higgs boson final state to probe the self-couplings.\footnote{Quadruple-Higgs boson production through longitudinal vector-boson scattering has
also been investigated in the SMEFT and HEFT frameworks~\cite{Delgado:2023ynh}.} Quadruple-Higgs boson
production has additional diagrammatic dependence on the quartic coupling and
can become comparatively sensitive to it in parts of the $(c_3,d_4)$ plane,
particularly for sizeable departures from the SM. This behaviour is
quantified through Eqs.~\eqref{eq:Rhhh} and~\eqref{eq:Rhhhh} and
Fig.~\ref{fig:xsecratio}, as discussed in the following section. Furthermore, in the final state that we examine
here, where each of the four Higgs bosons decays to $b\bar{b}$, the
higher-multiplicity irreducible QCD background is strongly suppressed by the
resolved jet requirements. For the restricted comparison of the tree-level
subprocesses $gg \rightarrow 6b$ and $gg \rightarrow 8b$, the cross section
falls by three orders of magnitude, from $\sim1$~pb to $\sim1$~fb at parton
level when requiring $p_{T,b} > 15$~GeV, $|\eta_b| < 3$, and
$\Delta R_{bb}>0.3$. By comparison, Fig.~\ref{fig:xsecratio} shows that, in
sizeable regions of
the $(c_3,d_4)$ plane away from the SM point, the ratio
$\sigma(gg\to hhhh\to8b)/\sigma(gg\to hhh\to6b)$ reaches values between
$0.01$ and $0.1$. In these regions, the quadruple-Higgs boson signal is therefore
suppressed by only one to two orders of magnitude relative to the
triple-Higgs boson signal.
These comparisons do not include reducible backgrounds
or mistags. Quadruple-Higgs boson production may therefore provide information on the
Higgs boson self-couplings that is complementary to lower-multiplicity $hh$ and
$hhh$ processes.

Beyond their sensitivity to the Higgs boson self-couplings, recent studies
have shown that multi-Higgs boson production can provide promising probes of
extended scalar sectors~\cite{Li:2026kqk,Roy:2026kud}. In particular, these
studies challenge the conventional expectation that new scalar states \(X\)
manifest themselves predominantly through decays into Higgs or vector-boson
pairs, or into fermions. Instead, regions of parameter space can arise in
which
\(
\Gamma\!\left(X\to hh,\,WW,\,ZZ,\,f\bar{f}\right)
\ll
\Gamma\!\left(X\to hhh\right),\,
\Gamma\!\left(X\to hhhh\right).
\)
Quadruple-Higgs boson production can therefore probe otherwise weakly-constrained
regions of extended scalar sectors. Related multi-Higgs boson production mechanisms
have also been investigated in
Refs.~\cite{Chen:2022vac,Lane:2024vur}.
More concretely, Ref.~\cite{Roy:2026kud} considered tuned scenarios in a two-real-singlet
extension with an integrated-out vector-like quark and in a one-singlet
extension with higher-dimensional operators. In those examples, an alignment
limit can suppress heavy-scalar decays to fermions and gauge bosons while
allowing enhanced multi-Higgs boson rates. Ref.~\cite{Li:2026kqk} considered a
specific $S(H^\dagger H)^3$ interaction involving a new TeV-scale scalar; the
large branching fractions quoted there, including values near $50\%$, apply
to that restricted interaction and normalization. Cascade production can
also make the $hhhh$ rate larger than the $hhh$ rate in parts of the complex
singlet model parameter space~\cite{Lane:2024vur}. These examples motivate a
phenomenological study of quadruple-Higgs boson production. We note that the
simplified topologies used below do not provide complete realizations of the cited models; they are used, instead, as illustrative examples.

The paper is organized as follows. In Sec.~\ref{sec:theory}, we discuss the
theoretical setup of Higgs boson self-coupling modifications and the
particular toy models of extended scalar sectors that we examine. In
Sec.~\ref{sec:pheno}, we describe the phenomenological analysis, including
Monte Carlo event generation, the simulation of detector effects through
jet-energy smearing and tagging, and the multivariate methods that we employ.
We then present the self-coupling and extended-scalar results. In the
self-coupling subsection, we also discuss possible contributions of
quadruple-Higgs boson production to the triple-Higgs boson signal region.
Conclusions and outlook are presented in Sec.~\ref{sec:conclusions}.
\ref{app:gsplit} describes the approximate simulation of triple
Higgs boson production in association with two $b$-jets, which forms part of
the eight-$b$-jet signal. \ref{app:classifier-observables} collects
the mathematical definitions of the multivariate classifier observables, and
\ref{app:self-coupling-score-diagnostics} presents supplementary classifier-score
results for the self-coupling analysis.

\section{Multi-Higgs Boson Production at Hadron Colliders}\label{sec:theory}
In what follows, we study two distinct scenarios. In the first, we explore the
possible constraints that can be derived at the HL-LHC on the plane of
modifications to the triple and quartic self-couplings using a combined
multi-Higgs boson hypothesis in the eight-$b$-jet final state. In the second
scenario, we investigate the possible constraints that can be obtained through
the resonant $hhhh\rightarrow8b$ final state on models that contain additional
scalar resonances. In this section, we present our theoretical setup for both
scenarios and motivate the use of quadruple-Higgs boson production in
constraining them.

\subsection{Self-coupling Modifications}

For the study of the Higgs boson self-couplings, we employ the phenomenological Lagrangian
\begin{equation}\label{eq:LgghhPheno}
  \mathcal{L} \supset - \frac{ M_h^2 } { 2 v } \Big( 1 + c_3 \Big) h^3 - \frac{ M_h^2 } { 8 v^2 }  \Big( 1  + d_4 \Big) h^4\;,
\end{equation}
where $h$ is the scalar field corresponding to the Higgs boson, $M_h \approx  125$~GeV is the Higgs boson mass~\cite{ATLAS:2012yve,CMS:2012qbp,ATLAS:2023owm,CMS:2024eka}, $v \approx 246$~GeV is the Higgs vacuum expectation value, and $c_3$ and $d_4$ parametrize deviations from the SM predictions for the triple and quartic self-couplings, respectively. In terms of the conventional coupling modifiers, $\kappa_3=1+c_3$ and $\kappa_4=1+d_4$.

\begin{figure}
    \centering
    \includegraphics[width=\linewidth]{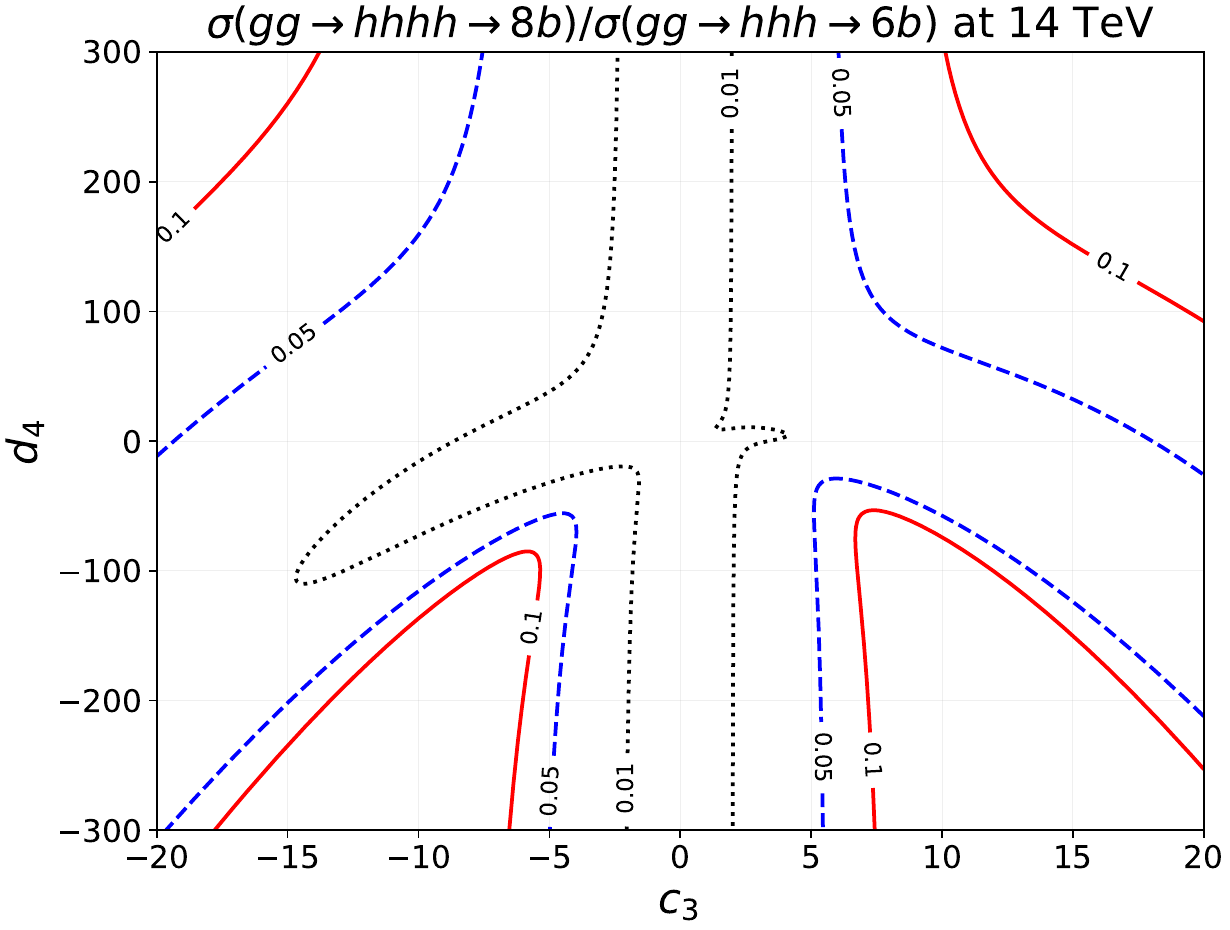}
    \caption{Contours of the ratio of LO cross sections of quadruple-Higgs boson production over triple-Higgs boson production to multi-$b$-quark final states at 14~TeV: $\sigma(gg\to hhhh \rightarrow 8b; c_3, d_4)/\sigma(gg\to hhh \rightarrow 6b; c_3, d_4)$, corresponding to the values $0.01$, $0.05$, and $0.1$.}
    \label{fig:xsecratio}
\end{figure}

Within the context of the above Lagrangian, for the loop-induced generation of
the signal samples, we employ \texttt{MadGraph5\_aMC@NLO} version
3.5.15~\cite{Alwall:2014hca,Hirschi:2015iia}, with a modified
\texttt{loop\_sm} model that incorporates the $c_3$ and $d_4$ modifications to
the triple and quartic self-couplings~\cite{Papaefstathiou:2023uum}, respectively. We perform leading-order (LO) fits of the 14~TeV cross sections for
$gg\rightarrow hhh$ and $gg\rightarrow hhhh$. For each process, the fit uses
the same 153-point coupling grid, spanning
\[
-30\leq c_3\leq30,\qquad -700\leq d_4\leq700.
\]
The PDF choices used by the different generators are
listed separately in Sec.~\ref{sec:event-generation}. Dividing by the LO SM
cross section in each case, we define the following ratio at a proton-proton centre-of-mass energy of $\sqrt{s}=14$~TeV, using
$gg\rightarrow nh$ as shorthand for the gluon-fusion contribution to the
hadronic rate:
\begin{equation}
R_{nh} \equiv 
\frac{\sigma(gg\to nh; c_3, d_4)}{\sigma(gg \rightarrow nh)_{\rm SM}}\;.
\end{equation}
For $gg \rightarrow hhh$, we then have
\begin{equation}\label{eq:Rhhh}
\begin{aligned}
R_{hhh}-1 \simeq{}&
 0.8709 c_3^2 - 0.8628 c_3 \\
&- 0.2589 c_3^3 - 0.1672 c_3d_4 \\
&- 0.09014 d_4 \\
&+ 0.04873 c_3^2d_4 \\
&+ 0.03947 c_3^4 + 0.01691 d_4^2\;.
\end{aligned}
\end{equation}
and, for $gg \rightarrow hhhh$,
\begin{equation}\label{eq:Rhhhh}
\begin{aligned}
R_{hhhh}-1 \simeq{}&
 0.5464 c_3^2 + 0.09615 c_3^4 \\
&+ 0.07839 c_3^3 + 0.03188 c_3 \\
&- 0.01833 c_3^5 + 0.003287 c_3^6 \\
&- 0.3621 d_4 - 0.1031 c_3d_4 \\
&- 0.1009 c_3^2d_4 \\
&- 0.03681 c_3^3d_4 \\
&+ 0.01296 c_3^4d_4 \\
&+ 0.04097 d_4^2 + 0.002992 c_3d_4^2 \\
&+ 0.01413 c_3^2d_4^2\;.
\end{aligned}
\end{equation}
The corresponding SM LO cross sections at 14~TeV are
\begin{equation}
\begin{aligned}
\sigma(gg \rightarrow hhh)_{\rm SM} \simeq{} & 0.0440~\mathrm{fb}\;,\\
\sigma(gg \rightarrow hhhh)_{\rm SM} \simeq{} & 1.24\times 10^{-4}~\mathrm{fb}\;,
\end{aligned}
\end{equation}
representing a reduction factor of $\sim 350$ going from three to four Higgs bosons. However, it is evident that $gg\rightarrow hhhh$ contains many more terms that probe modifications to the quartic self-coupling, as is expected from the larger number of diagrams. Nevertheless, both processes exhibit mild dependence on the quartic, particularly near SM-like trilinear couplings, $c_3 \sim 0$. To compare the dependence on the couplings, in Fig.~\ref{fig:xsecratio} we show contours on the $(c_3,d_4)$ plane of the cross-section ratio $\sigma(gg\to hhhh \rightarrow 8b; c_3, d_4)/\sigma(gg\to hhh\rightarrow 6b; c_3, d_4)$ at 14~TeV, including the branching fractions to the $8b$ and $6b$ final states. We show contours corresponding to the ratio values $0.01$, $0.05$, and $0.1$. Here and in what follows, we assume $\mathrm{BR}(h\rightarrow b\bar{b}) = 0.5824$.
There are regions of the $(c_3, d_4)$ plane where the cross section for the
quadruple-Higgs boson final state is between $5$ and $10\%$ of that for triple-Higgs boson production. Together with the suppression by approximately three orders of magnitude found in the restricted parton-level comparison of
$gg\rightarrow6b$ and $gg\rightarrow8b$ in Sec.~\ref{sec:intro}, this
motivates examining quadruple-Higgs boson production as a component of a combined
multi-Higgs boson constraint.

\subsection{Extended Scalar Sectors}

Motivated by Ref.~\cite{Roy:2026kud}, we consider two distinct extensions of the scalar sector containing either one or two additional resonances. Effective interactions between the scalars and gluons enable the production of these resonances at hadron colliders. For simplicity, we call the models the ``direct'' and ``cascade'' scenarios, according to the type of enhanced quadruple-Higgs boson production that can arise in each. For the direct scenario, the relevant terms in the phenomenological Lagrangian are
\begin{equation}\label{eq:Ldir}
\begin{aligned}
\mathcal{L}_\mathrm{dir} \supset & - \frac{1}{2} M_{h}^2 h^2 - \frac{1}{2} M_{S}^2 S^2 - \frac{M_h^2}{2v} h^3 - \frac{M_h^2}{8v^2} h^4 \\
&- \frac{c_5}{\Lambda} h^4 S + \frac{\alpha_s}{12\pi} \frac{C_{ggS}}{\Lambda} S G^a_{\mu\nu}  G^{a\mu\nu}\;.
\end{aligned}
\end{equation}
where $h$ is the SM-like Higgs boson and $S$ is the new scalar, with masses $M_h$ and $M_S$, respectively. The coefficient $c_5$ controls an effective operator coupling $S$ to four Higgs bosons, while $C_{ggS}$ controls the effective gluon-gluon-$S$ interaction. For the cascade scenario, we use
\begin{equation}\label{eq:Lcas}
\begin{aligned}
\mathcal{L}_\mathrm{cas} \supset  & - \frac{1}{2} M_1^2 h_1^2 - \frac{1}{2} M_2^2 h_2^2  - \frac{1}{2} M_3^2 h_3^2 \\
& - \frac{M_1^2}{2v} h_1^3 - \frac{M_1^2}{8v^2} h_1^4 \\
 &- \lambda_{112} h_1^2 h_2  - \lambda_{223} h_2^2 h_3 \\
& + \frac{\alpha_s}{12\pi} \frac{C_{gg3}}{\Lambda} h_3 G^a_{\mu\nu}  G^{a\mu\nu}\;,
\end{aligned}
\end{equation}
where $h_{1,2,3}$ are three physical scalar fields with masses $M_{1,2,3}$. The parameters $\lambda_{112}$ and $\lambda_{223}$ are the $h_1h_1h_2$ and $h_2h_2h_3$ scalar couplings, respectively, and $C_{gg3}$ controls the effective gluon-gluon-$h_3$ interaction. We identify $h_1$ as the SM-like Higgs boson and hence set $M_1=M_h$. In both scenarios, we assume that the triple and quartic Higgs self-couplings take their SM values.

The equations above display only the mass and interaction terms relevant to the
simulated topologies. Canonical kinetic terms are implicit and omitted for simplicity. Fermion,
gauge-boson, and additional scalar interactions are also omitted from the displayed
mass-basis description and are set to zero in the event-generation models.
The effective gluon interactions are normalized to resemble the SM Higgs boson
interaction in the $m_t \rightarrow \infty$ limit, but their normalization
does not enter the cross-section limits. Related, more complete
examples were constructed in Ref.~\cite{Roy:2026kud}, where
higher-dimensional operators were tuned towards an alignment limit with small
mixing between the observed Higgs boson and the new scalars. In such scenarios,
the partial widths can satisfy
\begin{equation}
\begin{aligned}
\Gamma(X\rightarrow WW,ZZ,f\bar{f}) &\ll \Gamma(X\rightarrow hhh)\;,\\
\Gamma(X\rightarrow WW,ZZ,f\bar{f}) &\ll \Gamma(X\rightarrow hhhh)\;.
\end{aligned}
\end{equation}
where $X$ denotes any of the new scalars. Higgs boson pair production can also be significant, although Ref.~\cite{Roy:2026kud} demonstrated that it can be suppressed through suitable choices of couplings.
\begin{figure*}[t]
  \centering
\includegraphics[height=0.2\textheight,keepaspectratio]{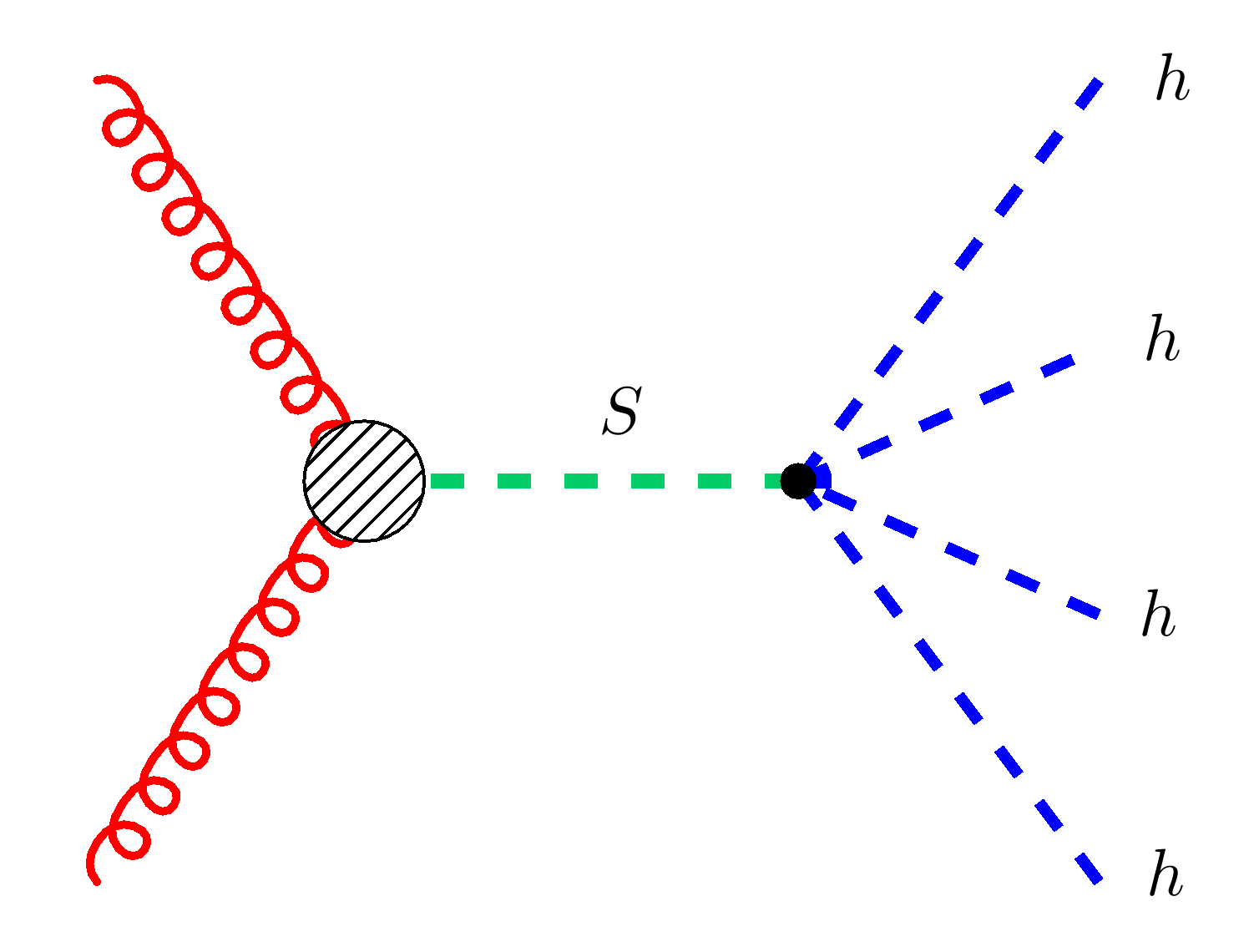}
  \qquad\qquad \includegraphics[height=0.2\textheight,keepaspectratio]{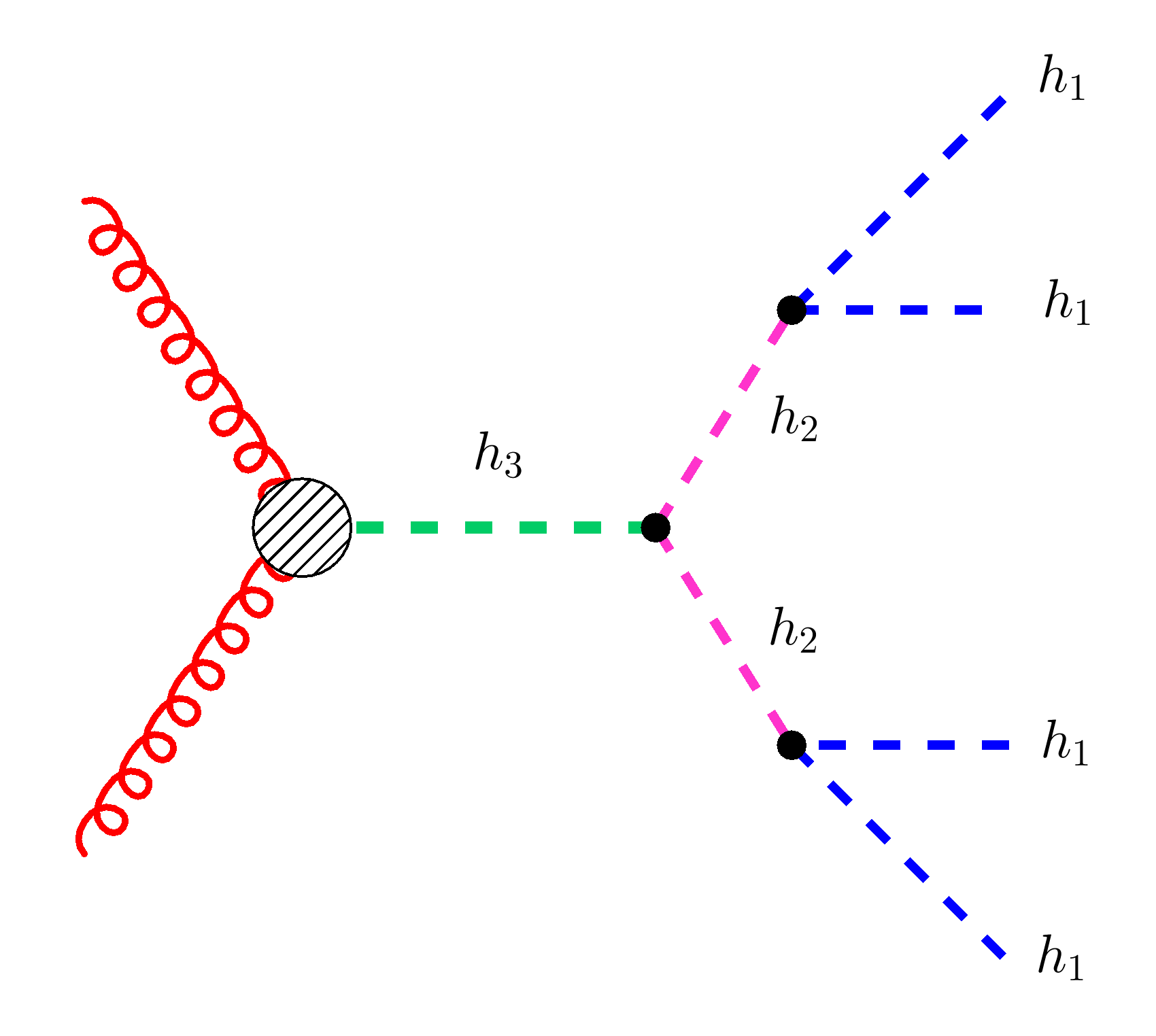}
  \caption{
    Feynman diagrams for quadruple-Higgs boson production in the two phenomenological models that we consider. Left: production of a heavy scalar $S$, followed by its direct decay into four Higgs bosons in the direct scenario. Right: the triple-resonant process in the cascade scenario, with production of $h_3$ and the subsequent decays $h_3\rightarrow h_2h_2\rightarrow 4h_1$. In both scenarios, $h$ and $h_1$ correspond to the SM-like Higgs boson. 
  }
  \label{fig:feynextended}
\end{figure*}
In the direct scenario, we examine the process in which a heavy scalar $S$, with $M_S > 4M_h$, is produced via the effective gluon interaction and decays resonantly into four SM-like Higgs bosons, as shown in the left panel of Fig.~\ref{fig:feynextended}. In the cascade scenario, we consider the triple-resonant cascade decay, where $h_3$ is produced via the effective gluon interaction and subsequently decays through $h_3 \rightarrow h_2 h_2$, with each $h_2$ decaying into two SM-like Higgs bosons, $h_1$. We require $M_3>2M_2$ and $M_2>2M_1$, which also imply $M_3>4M_1$.\footnote{The related nested decay
$h_1h_1\rightarrow4a_1\rightarrow8b$, requiring
$m_{h_1}>2m_{a_1}$, was recently studied in the
scNMSSM~\cite{Telba:2026vll}; it is distinct from the
quadruple-Higgs boson cascade considered here.} The corresponding Feynman diagram is shown in the right panel of Fig.~\ref{fig:feynextended}. We note that other diagrams can contribute to $gg\rightarrow hhhh$ in these models, particularly through additional scalar couplings. For simplicity, we omit them in this first study of quadruple-Higgs boson production, but we emphasize that they should be included in model-specific studies.

In this article, we employ these simplified mass-basis Lagrangians to study the
two displayed narrow-width topologies, without making a statement about other
signals present in a complete model. The corresponding rate parametrization
can be applied to a concrete model only when the narrow-width approximation
is adequate and the simulated kinematics describe that model. Broad
resonances, additional diagrams, or appreciable interference with each other
or with non-resonant $gg\rightarrow hhhh$ production require dedicated event
generation and cannot be inferred from the present templates.

\section{Phenomenological Analysis}\label{sec:pheno}

Here, we describe the collider setup, event generation, fiducial
selection, object definitions, and validation checks for each scenario considered.

\subsection{Event Generation}
\label{sec:event-generation}
The generation of the background processes contributing to the eight-$b$-jet final
state is one of the most challenging aspects of the present study.  For the
generation of the non-resonant $gg\rightarrow hhhh$ hard process in the
self-coupling modification scenario, we employ \texttt{MG5\_aMC}~3.5.15.  The
resonant signal samples are also generated with \texttt{MG5\_aMC} via custom UFO models~\cite{Darme:2023jdn}. For the generation of the multijet backgrounds, we employ a
patched version of the \texttt{Sherpa} Monte Carlo event generator with the
\texttt{Comix} matrix-element generator~\cite{Gleisberg:2008fv}, based on
commit \texttt{a7ba2c8b} of the 3.1.x development series~\cite{Sherpa:2024mfk}.

For clarity, the hard-process PDF choices are listed separately for signals
and backgrounds. The \texttt{MG5\_aMC} signal samples use its
\texttt{nn23lo1} choice, corresponding to
\texttt{NNPDF23\_lo\_as\_0130\_qed}. The
\texttt{Sherpa} $hh+4b$ signal uses
\texttt{NNPDF23\_nlo\_as\_0119}~\cite{Ball:2012cx}. All hard-process
background samples generated with \texttt{Sherpa} use the latter set as well.
For both signal and background LHE samples, the \texttt{HERWIG~7} shower,
LHE reader, multiparton interactions, and beam-remnant treatment use
\texttt{NNPDF23\_nlo\_as\_0119}. These sets were chosen for convenience rather than to assess PDF uncertainty; no PDF
variations are propagated to the results.
This \texttt{Sherpa} version has been modified to generate Les Houches event
(LHE) files with
the appropriate colour-flow topologies in the large-colour limit
($N_c\rightarrow\infty$).\footnote{In particular, the sampled
\texttt{Comix} colour flow is propagated to the LHE \texttt{ICOLUP} tags to
allow subsequent showering.  This version of \texttt{Sherpa} can be obtained
from the associated paper code repository~\cite{quadruplerepo}.}
Parton showering, hadronization, and multiparton interactions are simulated
using \texttt{HERWIG 7}~\cite{Bahr:2008pv,Gieseke:2011na,Arnold:2012fq,Bellm:2013hwb,Bellm:2015jjp,Bellm:2017bvx,Bellm:2019zci,Bewick:2023tfi,Bellm:2025pcw}, and the analysis is
implemented using the \texttt{HwSim} addon~\cite{hwsim}. The effects of pile-up are not considered. 

For the extended scalar sector analysis, we generate the direct process
$gg\rightarrow S\rightarrow hhhh$ and the triple-resonant cascade
$gg\rightarrow h_3\rightarrow h_2h_2\rightarrow 4h_1$ at
$\sqrt{s}=14$~TeV. The direct scan contains 42 mass
points in the range $525~\mathrm{GeV}\leq M_S\leq 5~\mathrm{TeV}$, with a
spacing of 25~GeV below 1~TeV, 100~GeV between 1 and 2~TeV, and 250~GeV above
2~TeV. The cascade scan contains 441 points in the physical region
$M_2>2m_{h_1}$ and $M_3>2M_2$, with
$275~\mathrm{GeV}\leq M_2\leq 2.4~\mathrm{TeV}$ and
$M_3\leq 5~\mathrm{TeV}$. The cascade plane is sampled more densely when either
resonance mass lies below 1~TeV and includes additional points close to the
$M_3=2M_2$ threshold. We fix the
widths of all new scalar resonances to 1~GeV. The resulting width-to-mass ratios satisfy $\Gamma/M\leq 3.6\times10^{-3}$ across the generated mass points. The narrow-width approximation therefore remains a central assumption of our analysis. In the direct case, only the
$ggS$ production and $Shhhh$ decay interactions are retained, while in the
cascade case only the $ggh_3$, $h_3h_2h_2$ and $h_2h_1h_1$ interactions, as well as the SM Higgs boson's self-couplings, are
nonzero. The generated cross sections are not used to normalize the signal,
and interference with the non-resonant SM $gg\rightarrow hhhh$ process is
neglected. The samples are subsequently processed with the same
\texttt{HERWIG 7} shower, hadronization and multiparton-interaction setup as the
remaining samples.

Throughout the analysis, we include the irreducible QCD background
$gg\rightarrow 8b$ and the reducible backgrounds
$gg\rightarrow 6b+2j_\ell$, $gg\rightarrow 6b+c\bar c$,
$gg\rightarrow 4b+4j_\ell$, $gg\rightarrow 4b+c\bar c+2j_\ell$, and
$gg\rightarrow 4b+4c$, where $j_\ell$ denotes a light quark or gluon.  We
also include the sample $gg\rightarrow h+6b$, with
$h\rightarrow b\bar b$, calculated in the infinite-top-mass limit (henceforth denoted by $m_t \rightarrow \infty$), and the tree-level process
$pp\rightarrow Z+6b$, with $Z\rightarrow b\bar b$.  Three gluon-initiated
$gg\rightarrow t\bar t+4b$ samples describe fully hadronic top decays and are
separated according to whether the four partons from the two $W$-boson decays
contain zero, one, or two charm quarks.  True $b$-jets are tagged with a probability $\epsilon_b = 0.85$. The charm and light jets receive mistag
probabilities $\epsilon_c=0.10$ and $\epsilon_j=0.01$, respectively. 

For the self-coupling inference, the signal hypothesis combines
non-resonant $gg\rightarrow hhhh$, the coupling-dependent
$gg\rightarrow hhhg$ process followed by $g\rightarrow b\bar b$, and
$gg\rightarrow hh+b\bar b b\bar b$.  The second contribution is
obtained using the forced-splitting approximation described and compared with
a large-$m_t$ reference sample in~\ref{app:gsplit}.\footnote{We emphasize that the process used in the limits is the full loop-induced process with the forced-splitting approximation applied, and not the one in the $m_t \rightarrow \infty$ limit.} We denote the third contribution by
$hh+4b$.  A simulation with full top-quark mass dependence is not
computationally tractable for the present study, so we calculate this process
in the $m_t\rightarrow\infty$ limit.  Even in this limit, event generation is
expensive; we therefore generate and shower a single SM sample and obtain the
$c_3$ dependence of its normalization from a weighted quadratic fit to
\texttt{Sherpa} integrations at $c_3=\{-20,-2,-1,0,20\}$:
\begin{equation}
\begin{aligned}
\sigma_{hh+4b}(c_3)\,[\mathrm{pb}]={}&
1.12\times10^{-5}-4.17\times10^{-6}c_3\\
&+1.15\times10^{-6}c_3^2\;.
\end{aligned}
\label{eq:hh4b-c3-fit}
\end{equation}
This is the raw LO generator cross section; the signal $K$-factor,
$\mathrm{BR}(h\rightarrow b\bar b)^2$, tagging probabilities and selection
efficiency are applied separately.  At every coupling point, we use the
classifier response and selection efficiency of the SM $hh+4b$ sample and
rescale only its normalization according to Eq.~\eqref{eq:hh4b-c3-fit}. This
should therefore be regarded as an indicative $m_t \rightarrow \infty$ estimate.

At each point in the $(c_3,d_4)$ plane, the relative normalization of all three
signal components is fixed by their predicted production cross sections. The
$hhh+b\bar b$ and $hh+4b$ samples are scored only after the classifier has been fixed. The classifier is trained using the SM $hhhh$
signal and the backgrounds. Neither additional signal is used in training or the background prediction. In
the extended scalar sector analysis, the non-resonant SM $gg\rightarrow hhhh$,
$gg \rightarrow hhh+b\bar{b}$ and $gg \rightarrow hh +4b$ samples are
included as additional backgrounds.  We
do not consider backgrounds containing hard leptons or missing transverse
momentum.
\begin{table}[t]
\centering
\begin{tabularx}{\columnwidth}{@{}lY@{}}
\toprule
\textbf{Object or pair} & \textbf{Generation-level requirement} \\
\midrule
Heavy-flavour partons &
All non-resonant $b$, $\bar{b}$, $c$, and $\bar{c}$ partons are required to
satisfy $p_T>15~\mathrm{GeV}$ and $|\eta|<3$. \\
Light jets &
All generated $j_\ell$ partons are required to satisfy
$p_T>15~\mathrm{GeV}$ and $|\eta|<3$. \\
Parton pairs &
Pairs of cut partons are required to satisfy $\Delta R>0.3$. \\
Resonance-decay exception &
For resonant $h\to b\bar{b}$ and $Z\to b\bar{b}$ decays, the daughter pair receives no $p_T$, $\eta$, or $\Delta R$
generation cuts. \\
\bottomrule
\end{tabularx}
\caption{Generation-level cuts used for the \texttt{Sherpa} background samples. The light-jet
container $j_\ell$ contains $d,\bar d,u,\bar u,s,\bar s,g$ and excludes charm.}\label{tab:generation_cuts}
\end{table}
\begin{table}[t]
\centering
\small
\begin{tabular}{p{0.64\columnwidth}r}
\toprule
\textbf{Sample} & \boldmath $\sigma_{\rm gen}\times{\rm BR}$ [fb] \\
\midrule
SM $gg \to hhhh \to 8b$ & $1.43\times 10^{-5}$ \\
SM $gg \to hhhg,\ g\to b\bar b,\ hhh\to 6b$ (forced splitting) &
$8.52\times 10^{-5}$ \\
SM $gg \to hh+4b,\ hh\to 4b$ $(m_t \rightarrow \infty)$ & $3.264\times 10^{-3}$ \\
\midrule
$gg \to 8b$ & $1.039$ \\
$gg \to 6b+2j_\ell$ & $680.8$ \\
$gg \to 6b+c\bar c$ & $5.238$ \\
$pp \to Z+6b,\ Z\to b\bar b$ & $3.378\times 10^{-2}$ \\
$gg \to 4b+4j_\ell$ & $2.086\times 10^{5}$ \\
$gg \to 4b+c\bar c+2j_\ell$ & $1570$ \\
$gg \to 4b+4c$ & $9.192$ \\
$gg \to h+6b,\ h\to b\bar b$ ($m_t\rightarrow \infty$) & $1.82\times 10^{-4}$ \\
$gg\rightarrow t\bar t+4b,\ 2c+2j_\ell$ & $0.3879$ \\
$gg\rightarrow t\bar t+4b,\ c+3j_\ell$ & $0.8097$ \\
$gg\rightarrow t\bar t+4b,\ 4j_\ell$ & $0.4144$ \\
\bottomrule
\end{tabular}
\caption{Generation-level cross sections after applying the relevant Higgs
boson and $Z$ boson branching fractions. No $K$-factors or tagging
probabilities are included. The $hhh+b\bar b$ entry uses the forced-splitting
approximation described in~\ref{app:gsplit}. The $hh+4b$
entry gives the event-sample normalization used to determine the classifier
response and efficiency; its normalization in the self-coupling analysis is
instead taken from the independent $c_3$-dependent fit in
Eq.~\eqref{eq:hh4b-c3-fit}. Integration uncertainties are considered only in
the secondary technical comparison described below.}\label{tab:sigma-gen-with-br}
\end{table}

We apply parton-level cuts as described in
Table~\ref{tab:generation_cuts}. Briefly, all partons that do not originate
from the decay of resonances (Higgs bosons or $Z$ bosons) are required to have
$p_T>15~\mathrm{GeV}$ and $|\eta|<3$ and to be separated by
$\Delta R>0.3$. A summary of generation-level cross sections, including the
three SM signal references $gg\rightarrow hhhh$,
$gg\rightarrow hhh+b\bar b$ and $gg\rightarrow hh+4b$, is shown in
Table~\ref{tab:sigma-gen-with-br}, with the appropriate branching ratios
applied where necessary, but without $K$-factors or tagging rates.

\subsection{Jet Energy Smearing}
\label{sec:jet-smearing}

In the self-coupling analysis, $b$-jets originating from Higgs bosons are expected to be resolved into separate jets, so we construct anti-$k_T$ $R=0.4$ jets (AK4 collection)
to investigate the $hhhh$ signal. In the extended scalar sectors analysis, when the parent scalar masses become large, $b$-jets from the boosted Higgs bosons merge into single ``fat'' jets. To capture this, in that analysis, we also cluster anti-$k_T$ $R=0.8$ jets (AK8 collection)~\cite{Cacciari:2008gp}. To account for the finite jet-energy resolution, we apply the CMS-like
energy-resolution parametrization used in Ref.~\cite{Conte:2012fm, Conte:2014zja, Conte:2018vmg, Araz:2020lnp,Fuks:2025gjv}. We first impose the raw $|\eta|<2.5$ acceptance. For
each retained jet, we draw one independent energy fluctuation according to
\begin{equation}
  E'=\max(10^{-6}~\mathrm{GeV},E+\Delta E),
  \qquad \Delta E\sim \mathcal{N}(0,\sigma_E),
  \label{eq:jet-energy-smearing}
\end{equation}
where the CMS-like energy resolution is
\begin{equation}
  \sigma_E^{\mathrm{CMS}}=
  \begin{cases}
    \sqrt{(0.05E)^2+(1.5)^2E}, & |\eta|\leq 3.0, \\
    \sqrt{(0.130E)^2+(2.7)^2E}, & 3.0<|\eta|\leq 5.0.
  \end{cases}
  \label{eq:cms-jet-energy-resolution}
\end{equation}
Here, both $E$ and $\sigma_E$ are expressed in GeV. Since all accepted jets
satisfy $|\eta|<2.5$, only the first branch of
Eq.~\eqref{eq:cms-jet-energy-resolution} contributes.

We then rescale the complete jet four-vector,
\begin{equation}
  p'^{\mu}=s p^{\mu},
  \qquad s=\frac{E'}{E},
  \label{eq:jet-four-vector-smearing}
\end{equation}
so that $p'_T=s p_T$ and $m'_j=s m_j$, while $\eta$ and $\phi$ are unchanged.
The strict transverse-momentum requirement is applied only after this
operation: $p'_T>20~\mathrm{GeV}$ for the AK4 self-coupling and resolved-resonance
collections, and $p'_T>300~\mathrm{GeV}$ for the boosted AK8 collection.
The upstream \texttt{HwSim} record stores jets down to
$p_T=10~\mathrm{GeV}$; hence the implemented upward-migration audit covers
the stored $10<p_T\leq20~\mathrm{GeV}$ population, while downward migrations
start above $20~\mathrm{GeV}$. Within that stored population, jets can migrate
across the analysis threshold in either direction. For an $R=0.8$
jet, the same scale factor is applied to its groomed and ungroomed
four-vectors (see the analysis description below for details). The resulting smeared four-vectors are used in the Higgs boson
reconstruction and multivariate observables, with no independent angular
smearing. We emphasize that the correlated four-vector scaling is an analysis-level extension of the
quoted energy-resolution model and is not a replacement for a full detector
simulation.

\subsection{Self-coupling Analysis}
\subsubsection{Separating Signal from Background}
\label{sec:multivariate-analysis}

After applying the generation requirements in
Table~\ref{tab:generation_cuts}, jet smearing and the resolved input selection,
we use \texttt{XGBoost}~\cite{Chen:2016btl} to distinguish quadruple-Higgs boson
events from the multijet backgrounds. The resolved selection requires eight
tagged AK4 jets with $p_T>20~\mathrm{GeV}$ after smearing and
$|\eta|<2.5$. The classifier assigns each event a score between zero and one.
Events with larger scores resemble the simulated $hhhh$ signal more closely,
while events with smaller scores resemble the backgrounds. The score ranks
events; it is not a calibrated probability that an event is signal.

We reconstruct four Higgs boson candidates from the eight selected jets. All
105 distinct pairings are examined. Within each pairing, the four candidates
are ordered by transverse momentum and compared with the target masses
$120$, $115$, $110$, and $105~\mathrm{GeV}$. We retain the pairing with the
smallest combined mass difference.\footnote{We find that these target masses provide better discrimination than using 125~GeV uniformly. In a future analysis, these values can be optimized to improve performance.} The 52 classifier inputs are listed in
Table~\ref{tab:xgboost-observables}. They describe the selected jets, the four
Higgs boson candidates, the quality of the pairing, and the overall event
kinematics. Their mathematical definitions are collected in~\ref{app:classifier-observables}. Event identifiers, event weights, and
the values of $(c_3,d_4)$ are not supplied to the classifier. The input
variables are fixed before any of the coupling-dependent signal samples are
examined.

\begin{table}[t]
\centering
\footnotesize
\begin{tabularx}{\columnwidth}{@{}l>{\centering\arraybackslash}p{0.09\columnwidth}Y@{}}
\toprule
\textbf{Information} & \textbf{No.} & \textbf{Observables} \\
\midrule
Selected jets & 8 & $p_T$ of each of the eight jets, ordered by decreasing
$p_T$. \\
Higgs boson candidates & 8 & The masses $m_{bb,a}$ and transverse momenta
$p_T(H_a)$ of the four candidates. \\
Pairing residuals & 5 & The four absolute differences between $m_{bb,a}$ and
the target masses $(120,115,110,105)~\mathrm{GeV}$, and their quadrature sum
$\chi_8$. \\
Angular separations & 10 & The six $\Delta R(H_a,H_b)$ separations and the four
$\Delta R_{bb}$ separations within the candidates. \\
Pairing ambiguity & 3 & The second-smallest $\chi_8$, its difference from the
best value, and the number among the 105 possible pairings with
$\chi_8<60~\mathrm{GeV}$. \\
Multi-Higgs boson masses & 10 & The six pairwise masses $m(H_a,H_b)$ and the four
three-candidate invariant masses. \\
Momentum sharing & 4 & The fraction $z_a$: the smaller of the two jet
transverse momenta divided by their sum, for each candidate. \\
Whole event & 4 & The eight-jet invariant mass $m_{8b}$, $H_T$ of the eight
jets, $p_T(4h)/m_{4h}$ and $|y(4h)|$. \\
\bottomrule
\end{tabularx}
\caption{The 52 reconstructed observables supplied to the classifier. See~\ref{app:classifier-observables} for their mathematical definitions. The
Higgs boson candidates are ordered by decreasing transverse momentum.}
\label{tab:xgboost-observables}
\end{table}

To train and validate the classifier, we use a dedicated SM
\(hhhh\to 8b\) sample, corresponding to
\((c_3,d_4)=(0,0)\). Once trained, the classifier is applied to statistically independent samples
on the same 153-point grid. For completeness, the grid includes the SM point
\((c_3,d_4)=(0,0)\); the sample evaluated at this point is statistically
independent of that used for training and validation.
Although the classifier is trained using \(hhhh\) events as signal, the
processes \(hhh+b\bar b\) and \(hh+4b\) populate the same selected eight-$b$-jet final state and therefore also contribute to the signal event
yield. The \(hhh+b\bar b\) contribution depends on both \(c_3\) and
\(d_4\), whereas, within the approximation adopted here, the \(hh+4b\)
contribution is taken to depend only on \(c_3\). This coupling dependence
is consistent with the polynomial dependence of the inclusive
double- and triple-Higgs boson signal strengths on the modified Higgs boson
self-couplings; see, for example, Eq.~(3.5) of Ref.~\cite{Haisch:2025pql}.\footnote{The small loop-induced dependence of double-Higgs boson production
on the quartic Higgs boson self-coupling is neglected in our treatment; see, e.g., Ref.~\cite{Haisch:2025pql}.}
In practice, the signal grid contains \(gg\to hhhh\) and
\(gg\to hhh+b\bar b\) samples at each of the \(153\) points in the
\((c_3,d_4)\) plane. The SM \(hh+4b\) sample is also evaluated with the
trained classifier. Its score distribution is kept fixed, while its
normalization is varied as a function of \(c_3\) according to
Eq.~\eqref{eq:hh4b-c3-fit}. The predicted signal yield in each classifier
bin is then obtained by adding the contributions from all three
processes.

We use only
$hhhh$ to define the signal class because quadruple-Higgs boson production is the
primary process under study. This also keeps the classifier independent of an
assumed mixture of the three signal processes. Their different coupling
dependences are introduced later through their physical score distributions
and rates.

The weights used to train the classifier are different from those used to
predict event yields. During training, all weights are non-negative. The
relative contributions of the background processes are retained, but the
signal class and the combined background class are given equal total weight.
This prevents the much larger background sample from dominating the training.
The classifier settings are chosen with the validation samples before the
physical coupling grid is evaluated.\footnote{We compare six predefined
configurations: the baseline choice of 300 trees, maximum depth three,
learning rate 0.05, and row and feature subsampling fractions of 0.9;
maximum depths two and four; a minimum child weight of five; an $L_2$
regularization parameter of five; and 500 trees with learning rate 0.03. We
rank them by the average expected SM $hhhh$ sensitivity obtained with the five-bin candidate score distributions, retaining the baseline if
it is within 1\% of the best result. This selects the baseline settings with
the $L_2$ regularization parameter increased from one to five. No
coupling-dependent sample enters this comparison.}

For the rate prediction, the physical weight of event $i$ from process $p$ is
\begin{equation}
  w^{\mathrm{phys}}_{i,p}
  =\mathcal{L}\,\sigma_p F_p\,
   \frac{w_{i,p}}{W_p^{\mathrm{input}}},
  \label{eq:self-coupling-physical-weight}
\end{equation}
where $\mathcal{L}$ is the integrated luminosity, $\sigma_p$ is the production
cross section, and $F_p$ contains the relevant branching fractions,
tagging probabilities, and higher-order correction factor (i.e.\ the $K$-factor). Here $w_{i,p}$
is the Monte Carlo event weight and $W_p^{\mathrm{input}}$ is the sum of the
weights before the analysis selection. We use
$\mathcal{L}=3000~\mathrm{fb}^{-1}$,
$K_{\mathrm{s}}=K_{\mathrm{b}}=2$,
$\mathrm{BR}(h\rightarrow b\bar b)=0.5824$~\cite{LHCHiggsCrossSectionWorkingGroup:2016ypw},
$\epsilon_b=0.85$, $\epsilon_c=0.10$, and $\epsilon_j=0.01$. Thus, for
example, the non-resonant $hhhh$ signal has
$F_{\mathrm{s}}=K_{\mathrm{s}}
\left[\mathrm{BR}(h\rightarrow b\bar b)\right]^4\epsilon_b^8$. For the $hhh+b\bar b$ and $hh+4b$ samples, the factors $F_p$ depend on the third and second powers of $\mathrm{BR}(h\rightarrow b\bar b)$, respectively. The tagging factor obeys the general structure $\epsilon_b^{n_b}\epsilon_c^{n_c}\epsilon_j^{n_j}$. Tagging probabilities are
therefore applied once. Signed generator weights are retained when calculating
selection efficiencies, score distributions, and event yields.

To avoid evaluating the classifier on events used for its training, we divide
the dedicated training signal sample and each background sample into five
subsamples with fixed numerical labels that remain unchanged throughout the
analysis. To exploit all generated events while maximizing the robustness of
the statistical analysis, we ``rotate'' the roles of these subsamples. We
perform five rotations, resulting in five independently trained classifiers.
In each rotation, three of the five subsamples are used for training, one for
validation, and the remaining one for evaluation (the ``held-out''
subsample). The validation subsample is used to monitor and optimize the
training procedure, whereas the held-out subsample is used to obtain the
final classifier output. The rotations are arranged such that every
subsample serves as the held-out subsample exactly once. Consequently, every
event is evaluated exactly once by a classifier that was not trained on that
event.

\subsubsection{From Classifier Scores to Confidence Regions}
\label{sec:self-coupling-likelihood}
Rather than imposing a single cut on the classifier output, we use its
binned distribution (the ``classifier shape'') in the statistical analysis.
We convert the classifier scores into confidence regions through the
following six steps.

\medskip
\noindent\textbf{Step 1: Determine the score-bin boundaries.}
For each rotation, we apply the trained classifier to the background events
in the validation subsample. We use the resulting weighted score distribution
to determine the numerical boundaries corresponding to the following
background-percentile intervals:
\begin{equation}
\begin{aligned}
&[0,0.50),\qquad [0.50,0.75),\qquad [0.75,0.90),\\
&[0.90,0.97),\qquad [0.97,1]\;.
\end{aligned}
\label{eq:percentiles}
\end{equation}
By construction, these intervals contain approximately \(50\%\), \(25\%\),
\(15\%\), \(7\%\), and \(3\%\) of the weighted validation background,
respectively. They therefore represent percentiles of the background score
distribution rather than fixed numerical \texttt{XGBoost}-score intervals.

\medskip
\noindent\textbf{Step 2: Apply the boundaries to the held-out samples.}
We determine the numerical score boundaries separately for each rotation and
apply them only to the corresponding held-out subsample. Because the five
classifiers are trained independently, their numerical boundaries need not be
identical; see Fig.~\ref{fig:c3d4-five-classifier-scores}. Once determined,
however, the boundaries remain fixed and are applied unchanged to all
coupling-dependent samples throughout the \((c_3,d_4)\) plane.

\medskip
\noindent\textbf{Step 3: Combine the five rotations.}
We add the event yields in corresponding percentile intervals across the five
rotations: the yields in the first interval are combined into one bin, those
in the second interval into another, and so forth. This produces a single
five-bin event-count distribution in which every event is evaluated exactly
once by a classifier that was not trained on it.

\noindent
\ref{app:self-coupling-score-diagnostics} shows the five individual
classifier-score distributions and their combination. These distributions
verify that the boundaries obtained from the validation events produce the
intended background fractions when applied to the held-out samples. The
individual distributions shown there are normalized to unit area to
facilitate comparison of their shapes, whereas the likelihood uses the
physical event yields in the five combined bins.

\medskip
\noindent\textbf{Step 4: Construct the coupling-dependent signal prediction.}
At a fixed point \((c_3,d_4)\), the predicted signal in score bin \(i\) is
\begin{equation}
\begin{aligned}
S_i(c_3,d_4)={}&S_i^{hhhh}(c_3,d_4)
 +S_i^{hhh b\bar b}(c_3,d_4)\\
&+\frac{\sigma_{hh+4b}(c_3)}{\sigma_{hh+4b}(0)}
 S_i^{hh+4b}(0,0)\;.
\end{aligned}
\label{eq:self-coupling-count-model}
\end{equation}
Here, \(S_i^{hh+4b}(0,0)\) is the SM score-bin yield obtained from the
score shape and selection efficiency of the \(hh+4b\) sample and normalized
using \(\sigma_{hh+4b}(0)\) from Eq.~\eqref{eq:hh4b-c3-fit}. The cross-section
ratio therefore supplies only the relative \(c_3\) dependence.

We obtain the \(hhhh\) and \(hhh b\bar b\) templates directly from the
physical samples generated at all 153 points in the \((c_3,d_4)\) plane. For
\(hh+4b\), we retain the score distribution and selection efficiency of the
SM sample while rescaling its normalization according to
Eq.~\eqref{eq:hh4b-c3-fit}. The three contributions thus retain their physical
normalizations and different coupling dependences. The resulting constraint
therefore applies to the combined multi-Higgs boson signal rather than to
\(hhhh\) alone.

\begin{table*}[t]
\centering
\footnotesize
\setlength{\tabcolsep}{3.0pt}
\renewcommand{\arraystretch}{1.06}
\setlength{\dashlinedash}{2pt}
\setlength{\dashlinegap}{2pt}
\begin{tabularx}{\textwidth}{@{}Yrrrrrr@{}}
\toprule
\textbf{Sample} &
\boldmath $c_3$ &
\boldmath $d_4$ &
\shortstack{\boldmath $\sigma_{\rm input}^{8\,\mathrm{AK4}}$\\$[\mathrm{fb}]$} &
\shortstack{\boldmath $N_{\rm input}^{8\,\mathrm{AK4}}$\\\textbf{expected}} &
\shortstack{\boldmath $\sigma_{\rm XGB}$\\$[\mathrm{fb}]$} &
\shortstack{\boldmath $N_{\rm XGB}$\\\textbf{expected}} \\
\midrule
\multicolumn{7}{@{}l}{\emph{Standard Model signal references}} \\
SM $gg\to hhhh$ & $0$ & $0$ & $6.743\times10^{-7}$ &
$2.023\times10^{-3}$ & $2.137\times10^{-7}$ & $6.411\times10^{-4}$ \\
SM $gg\to hhh+b\bar b$ & $0$ & $0$ & $2.490\times10^{-6}$ &
$7.470\times10^{-3}$ & $5.220\times10^{-7}$ & $1.566\times10^{-3}$ \\
SM $gg\to hh+4b$ & $0$ & $0$ & $1.280\times10^{-4}$ &
$0.3841$ & $1.749\times10^{-5}$ & $0.05247$ \\
\textbf{Total SM signal} & $0$ & $0$ & $\mathbf{1.312\times10^{-4}}$ &
\textbf{0.3936} & $\mathbf{1.822\times10^{-5}}$ & \textbf{0.05467} \\
\midrule
\multicolumn{7}{@{}l}{\emph{Representative non-SM signal points}} \\
$gg\to hhhh$ & $0$ & $-200$ & $1.159\times10^{-3}$ & $3.476$ &
$3.765\times10^{-4}$ & $1.129$ \\
$gg\to hhh+b\bar b$ & $0$ & $-200$ & $1.546\times10^{-3}$ & $4.638$ &
$2.677\times10^{-4}$ & $0.8032$ \\
\hdashline
$gg\to hhhh$ & $0$ & $200$ & $1.068\times10^{-3}$ & $3.205$ &
$3.234\times10^{-4}$ & $0.9701$ \\
$gg\to hhh+b\bar b$ & $0$ & $200$ & $1.473\times10^{-3}$ & $4.419$ &
$2.939\times10^{-4}$ & $0.8818$ \\
\hdashline
$gg\to hhhh$ & $-9$ & $0$ & $1.838\times10^{-3}$ & $5.513$ &
$4.774\times10^{-4}$ & $1.432$ \\
$gg\to hhh+b\bar b$ & $-9$ & $0$ & $9.122\times10^{-4}$ & $2.737$ &
$1.346\times10^{-4}$ & $0.4038$ \\
\hdashline
$gg\to hhhh$ & $12$ & $0$ & $4.030\times10^{-3}$ & $12.09$ &
$1.029\times10^{-3}$ & $3.086$ \\
$gg\to hhh+b\bar b$ & $12$ & $0$ & $7.905\times10^{-4}$ & $2.372$ &
$1.114\times10^{-4}$ & $0.3342$ \\
\hdashline
$gg\to hhhh$ & $7.5$ & $50$ & $2.036\times10^{-3}$ & $6.107$ &
$6.061\times10^{-4}$ & $1.818$ \\
$gg\to hhh+b\bar b$ & $7.5$ & $50$ & $3.664\times10^{-4}$ & $1.099$ &
$5.928\times10^{-5}$ & $0.1778$ \\
\hdashline
$gg\to hhhh$ & $-1.5$ & $150$ & $9.848\times10^{-4}$ & $2.954$ &
$2.960\times10^{-4}$ & $0.8879$ \\
$gg\to hhh+b\bar b$ & $-1.5$ & $150$ & $8.752\times10^{-4}$ & $2.626$ &
$1.565\times10^{-4}$ & $0.4694$ \\
\hdashline
$gg\to hhhh$ & $-6$ & $-100$ & $2.407\times10^{-3}$ & $7.221$ &
$7.464\times10^{-4}$ & $2.239$ \\
$gg\to hhh+b\bar b$ & $-6$ & $-100$ & $1.236\times10^{-4}$ & $0.3707$ &
$2.289\times10^{-5}$ & $0.06866$ \\
\hdashline
$gg\to hhhh$ & $6$ & $-100$ & $3.305\times10^{-3}$ & $9.915$ &
$9.988\times10^{-4}$ & $2.996$ \\
$gg\to hhh+b\bar b$ & $6$ & $-100$ & $3.359\times10^{-4}$ & $1.008$ &
$5.235\times10^{-5}$ & $0.1570$ \\
\addlinespace
\multicolumn{7}{@{}l}{\emph{Background samples}} \\
$gg\to8b$ & -- & -- & $4.884\times10^{-3}$ & $14.65$ & $1.804\times10^{-4}$ & $0.5411$ \\
$gg\to6b+2j_\ell$ & -- & -- & $1.422\times10^{-3}$ & $4.267$ & $2.096\times10^{-5}$ & $0.06287$ \\
$gg\to6b+c\bar c$ & -- & -- & $4.069\times10^{-4}$ & $1.221$ & $8.258\times10^{-6}$ & $0.02477$ \\
$pp\to Z+6b$, $Z\to b\bar b$ & -- & -- & $3.509\times10^{-4}$ & $1.053$ & $1.564\times10^{-5}$ & $0.04692$ \\
$gg\to4b+4j_\ell$ & -- & -- & $1.253\times10^{-4}$ & $0.3758$ & $2.779\times10^{-6}$ & $0.008338$ \\
$gg\to4b+c\bar c+2j_\ell$ & -- & -- & $8.320\times10^{-5}$ & $0.2496$ & $6.464\times10^{-7}$ & $0.001939$ \\
$gg\to4b+4c$ & -- & -- & $1.138\times10^{-5}$ & $0.03413$ & $3.779\times10^{-7}$ & $0.001134$ \\
$gg\to h+6b$, $h\to b\bar b$ & -- & -- & $2.233\times10^{-6}$ & $0.006698$ & $1.849\times10^{-7}$ & $5.548\times10^{-4}$ \\
$t\bar t+4b$, $2c+2j$ & -- & -- & $1.583\times10^{-7}$ & $4.748\times10^{-4}$ & $1.404\times10^{-8}$ & $4.211\times10^{-5}$ \\
$t\bar t+4b$, $c+3j$ & -- & -- & $4.091\times10^{-8}$ & $1.227\times10^{-4}$ & $3.900\times10^{-9}$ & $1.170\times10^{-5}$ \\
$t\bar t+4b+4j$ & -- & -- & $2.418\times10^{-9}$ & $7.255\times10^{-6}$ & $2.406\times10^{-10}$ & $7.219\times10^{-7}$ \\
\textbf{Total background} & -- & -- & $\mathbf{7.286\times10^{-3}}$ &
\textbf{21.86} & $\mathbf{2.292\times10^{-4}}$ & \textbf{0.6877} \\
\midrule
\textbf{SM signal plus background} & $0$ & $0$ & $\mathbf{7.417\times10^{-3}}$ &
\textbf{22.25} & $\mathbf{2.474\times10^{-4}}$ & \textbf{0.7423} \\
\bottomrule
\end{tabularx}
\caption{Expected cross sections and event counts at
$\sqrt{s}=14~\mathrm{TeV}$ and $\mathcal{L}=3000~\mathrm{fb}^{-1}$. The input
selection requires eight separately tagged AK4 jets with
$p_T>20~\mathrm{GeV}$ after smearing and $|\eta|<2.5$. Thus,
$\sigma_{\rm input}^{8\,\mathrm{AK4}}$ is the effective cross section after
this resolved selection, including the branching fractions, $K$-factor and
fixed tagging or mistagging probabilities. It is not the hard-process
generator cross section, and
$N_{\rm input}^{8\,\mathrm{AK4}}=\mathcal{L}
\sigma_{\rm input}^{8\,\mathrm{AK4}}$. No AK8 candidates are used. The
\texttt{XGBoost} columns show the fifth, highest-score bin after the matching
held-out bins have been combined across the five rotations. They illustrate
the classifier response and do not define a score cut; all five bins enter the
likelihood. The generator cross sections are the same as those used in
Table~\ref{tab:resonance-score-yields}, while the effective input cross
sections differ because that analysis uses a hybrid AK4+AK8 selection. Both
tables use
$\sigma_{\rm LHE}(gg\to4b+c\bar c+2j_\ell)=2.75178~\mathrm{pb}$ before the common
$K$-factor of two. The fitted $hh+4b$ contribution is displayed only at the SM
point but is included at every tested coupling point.}
\label{tab:self-coupling-score-yields}
\end{table*}

\medskip
\noindent\textbf{Step 5: Construct the likelihood.}
We treat the five combined percentile intervals as separate event counts,
labelled by \(i=1,\ldots,5\), and denote the expected background in bin \(i\)
by \(B_i\). To calculate the median expected sensitivity, we construct an SM
signal-plus-background \emph{Asimov data set}. This artificial data set
contains no random fluctuations: the count in each bin is set equal to its
SM expectation,
\begin{equation}
n_i=B_i+S_i(0,0)\;.
\end{equation}
The SM signal in this expression is obtained from the physical grid samples
rather than from the separate sample used to train the classifier. At a
tested point, the expected count in the same bin is
\begin{equation}
\nu_i(c_3,d_4)=B_i+S_i(c_3,d_4)\;.
\end{equation}
These expected event counts may be fractional.

Each bin contributes a separate Poisson factor, giving the likelihood
\begin{equation}
\lambda(c_3,d_4)
=\prod_{i=1}^{5}
\operatorname{Pois}\!\left[n_i\,\middle|\,\nu_i(c_3,d_4)\right]\;,
\label{eq:self-coupling-poisson-likelihood}
\end{equation}
where
\begin{equation}
\operatorname{Pois}\!\left[n_i\,\middle|\,\nu_i(c_3,d_4)\right]
=
\frac{\bigl[\nu_i(c_3,d_4)\bigr]^{n_i}}{\Gamma(n_i+1)}
e^{-\nu_i(c_3,d_4)}\;.
\label{eq:poisson}
\end{equation}
The factor $\Gamma(n_i+1)$ analytically continues the factorial to fractional
Asimov counts. It is independent of the tested parameters and cancels in the
likelihood ratio.
The product compares the predicted and SM event counts separately in all five
score regions. It therefore uses both the total event rate and the
distribution of events across the classifier score. No score bin is
discarded and no threshold cut is applied.

\medskip
\noindent\textbf{Step 6: Calculate the test statistic and confidence regions.}
We compare each tested point with the SM point using
\begin{equation}
\begin{aligned}
q(c_3,d_4)
&=-2\ln\frac{\lambda(c_3,d_4)}{\lambda(0,0)}\\
&=2\sum_{i=1}^{5}\left[
\nu_i(c_3,d_4)-n_i
+n_i\ln\frac{n_i}{\nu_i(c_3,d_4)}
\right]\;.
\end{aligned}
\label{eq:self-coupling-poisson-q}
\end{equation}
The SM point satisfies \(q(0,0)=0\). Using the standard asymptotic
approximation, the joint \(68\%\) and \(95\%\) confidence contours for the two
couplings are defined by \(q=2.30\) and \(q=5.991\), respectively. When one
coupling is fixed to its SM value, the remaining scan contains one parameter
and the \(95\%\) interval is defined by \(q=3.841\).
Some bins contain sub-unit expected counts; the asymptotic contours are
therefore interpreted as approximate statistics-only projections. Validation
with pseudoexperiments is beyond the scope of the present study.

We evaluate Eq.~\eqref{eq:self-coupling-poisson-q} at every physical grid
point. We do not use interpolation to construct signal templates or event
counts; a smooth Clough--Tocher interpolation of the calculated \(q\) values
is used only to draw the final \(95\%\) contour.

The nominal calculation includes only Poisson counting statistics and does
not incorporate systematic uncertainties. Since the tree-level background
cross sections are expected to carry a large uncertainty, we repeat the
calculation using \(B_i/4\) and \(4B_i\), applying the same rescaling to both
the Asimov data and the tested expectation. These calculations are
illustrative stress tests rather than a systematic-uncertainty band.

We do not include detector, tagging, PDF, scale, finite-simulation, or
background-shape uncertainties. We also omit uncertainties associated with
the \(hhh+b\bar b\) forced-splitting approximation, the \(hh+4b\)
normalization fit, and the \(m_t\to\infty\) approximation. The resulting
confidence regions should therefore be understood as approximate
statistics-only projections.

Table~\ref{tab:self-coupling-score-yields} gives the expected input rates and
the yield in the highest-score bin for the SM signals, representative non-SM
coupling points, and the backgrounds. This one-bin summary, corresponding to
the rightmost bar in
Fig.~\ref{fig:c3d4-combined-percentile-background}, illustrates the
classifier response; all five bins enter the limit calculation.

\subsubsection{Self-Coupling Constraints from Combined Multi-Higgs Boson Production}
\label{sec:self-coupling-results}

At the SM point, $(c_3,d_4)=(0,0)$, the resolved eight-tag input selection yields
0.394 expected signal events and 21.9 background events across the five
classifier-score bins. The largest signal contribution at the SM point is the
$hh+4b$ process, although it becomes less important for larger self-coupling
modifications.

The simultaneous two-dimensional $(c_3, d_4)$ 68\% region projects onto
$-9.1<c_3<10.1$ and $-163<d_4<168$, while the simultaneous 95\% region
projects onto
\begin{equation}
 -10.3<c_3<11.2, \qquad -222<d_4<234.
 \label{eq:self-coupling-projected-limits}
\end{equation}

For completeness, we also perform the two fixed-coupling scans using the same
pointwise likelihood. Applying the one-parameter 95\%
threshold $q=3.841$ gives
\begin{equation}
 \begin{aligned}
  c_3=0:&\quad -191<d_4<197,\\
  d_4=0:&\quad -7.6<c_3<10.3.
 \end{aligned}
 \label{eq:self-coupling-fixed-limits}
\end{equation}

\begin{figure*}[t]
  \centering
  \includegraphics[width=0.94\textwidth]{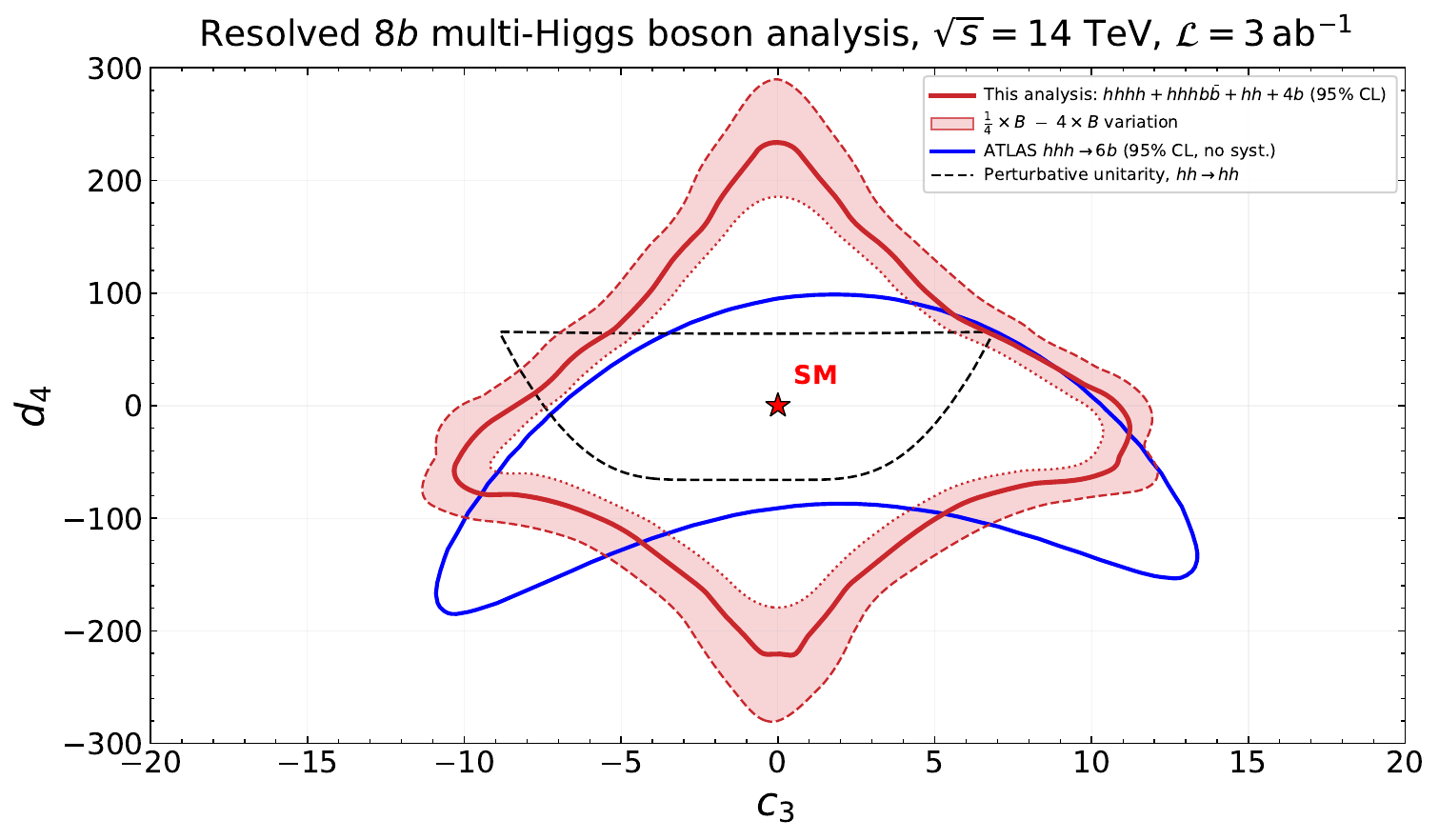}
  \caption{Expected simultaneous constraints in the $(c_3,d_4)$ plane from
  the binned \texttt{XGBoost}-score Poisson likelihood. The solid red curve is the
  nominal statistics-only 95\% contour. The pale-red region and its dotted and
  dashed boundaries show the $B/4$ and $4B$ background-level illustrative stress tests; it
  is not a systematic-uncertainty band. The blue curve is the simultaneous
  no-systematics $hhh\to6b$ projection from Ref.~\cite{ATLAS:2025cae},
  translated using $c_3=\kappa_3-1$ and $d_4=\kappa_4-1$. The black dashed
  curve indicates the perturbative-unitarity boundary from $hh \rightarrow hh$ scattering, and the star marks the
  SM point. This self-coupling result uses the resolved eight-AK4-jet input
  selection specified in Table~\ref{tab:self-coupling-score-yields}.}
  \label{fig:c3d4-simple-poisson-contour}
\end{figure*}

Fig.~\ref{fig:c3d4-simple-poisson-contour} shows the expected simultaneous
95\% confidence contour from the nominal statistics-only likelihood. The
$B/4$ and $4B$ calculations delimit the pale-red stress-test envelope obtained by rescaling the background normalization downwards and upwards by a factor of four. Their separation illustrates the sensitivity of the contour to the assumed background level; it is not an uncertainty estimate or band. The perturbative-unitarity boundary from $hh\rightarrow hh$ scattering and the SM point are shown for reference~\cite{DiLuzio:2017tfn, Liu:2018peg, Stylianou:2023tgg,Fuks:2025gjv}.

For context, we overlay the simultaneous no-systematics HL-LHC projection for
$hhh\to6b$ from Ref.~\cite{ATLAS:2025cae}, after the translation
$c_3=\kappa_3-1$ and $d_4=\kappa_4-1$. The coordinate extrema of the black
no-systematics contour in Fig.~7 of Ref.~\cite{ATLAS:2025cae}, read from the
published figure, are approximately $-11\lesssim c_3\lesssim13$ and
$-185\lesssim d_4\lesssim99$. These values are projections of a digitized
simultaneous two-parameter contour; they are not one-dimensional intervals
quoted by ATLAS. On this like-for-like basis, the two analyses have comparable
projected reach in $c_3$, while the ATLAS contour is narrower in $d_4$.

ATLAS separately quotes the no-systematics one-parameter 95\% interval
$-79<\kappa_4<90$ for fixed $\kappa_3=1$, corresponding to
$-80<d_4<89$. This should be compared with our fixed-$c_3$ interval in
Eq.~\eqref{eq:self-coupling-fixed-limits}, rather than with the projection of
our simultaneous contour. Ref.~\cite{ATLAS:2025cae} does not quote the
corresponding no-systematics HL-LHC one-parameter interval in $\kappa_3$ for
fixed $\kappa_4=1$, so we do not infer one from its two-dimensional contour.
The comparison remains between complementary analyses: the ATLAS projection
contains only the $hhh$ signal, whereas our result uses the combined
multi-Higgs boson hypothesis in Eq.~\eqref{eq:self-coupling-count-model}.

\subsection{Contribution of Quadruple-Higgs Boson Production to the Triple-Higgs Boson Production Signal Region}

Although conventionally denoted the ``$6b$'' signal region, the ATLAS
$hhh\to6b$ analysis requires \emph{at least} six $b$-tagged jets, rather than
exactly six~\cite{ATLAS:2024xcs}. The event categories contain exactly four,
exactly five, and at least six tagged jets, respectively. In events with more
than six tags, the six highest-$p_T$ tagged jets are used to reconstruct the
three Higgs boson candidates. The HL-LHC projection retains this Run~2 analysis
definition~\cite{ATLAS:2025cae}. Consequently, $hhhh\to8b$ events with seven or
eight tagged jets enter the ATLAS signal region directly.

This inclusive definition substantially increases the potential contribution
from quadruple-Higgs boson production. Based solely on the tagging efficiency,
the probability to tag exactly $k$ out of the eight $b$-jets within the $b$-tagging transverse momentum and pseudorapidity thresholds is approximately
\begin{equation}
P(k)=\binom{8}{k}(\epsilon_b)^k(1 - \epsilon_b)^{8-k}\;,
\end{equation}
yielding probabilities of approximately $23.8\%$, $38.5\%$, and $27.2\%$ for
exactly six, seven, and eight tags, respectively, for $\epsilon_b=0.85$. Thus,
before kinematic acceptance effects,
\begin{equation}
 P(N_b^{\rm tag}\geq6)=\sum_{k=6}^{8}P(k)\simeq89.5\%
\end{equation}
for $hhhh\to8b$. For $hhh\to6b$, the corresponding probability is
$P(N_b^{\rm tag}\geq6)=\epsilon_b^6\simeq37.7\%$ when only the six
$b$-jets from Higgs boson decays are present.

We quantify the relative quadruple-Higgs boson contribution through
a loose fiducial analysis across the
$(c_3,d_4)$ plane. More concretely, we use the ratio
\begin{equation}
R_{\rm fid} =
\frac{\sigma_{\rm fid}(gg\to hhhh)}
{\sigma_{\rm fid}(gg\to hhh)
+\sigma_{\rm fid}(gg\to hhh+b\bar b)}\;.
\end{equation}
The explicit \(hhh+b\bar b\) contribution is obtained from full-loop
\(gg\to hhhg\) events followed by a forced \(g\to b\bar b\) splitting.
This contribution is added without matching to the inclusive \(hhh\)
sample; in particular, overlap with heavy flavour generated by its
parton shower is not removed. The double counting increases
the denominator of \(R_{\rm fid}\) and therefore leads to a
conservative estimate of the relative \(hhhh\) contribution. 
The term ``fiducial''
refers to the following phase-space region, defined by these event-selection
requirements: we use the same CMS-like 
jet-energy smearing as in the quadruple-Higgs boson analysis and a $b$-tagging efficiency of
$\epsilon_b = 0.85$, without mistags. We require at least six tagged $b$-jets with
$p_T>20$~GeV and $|\eta|<2.5$ for both $hhhh$ and $hhh$.

\begin{figure}[t]
  \centering
  \includegraphics[width=\columnwidth]{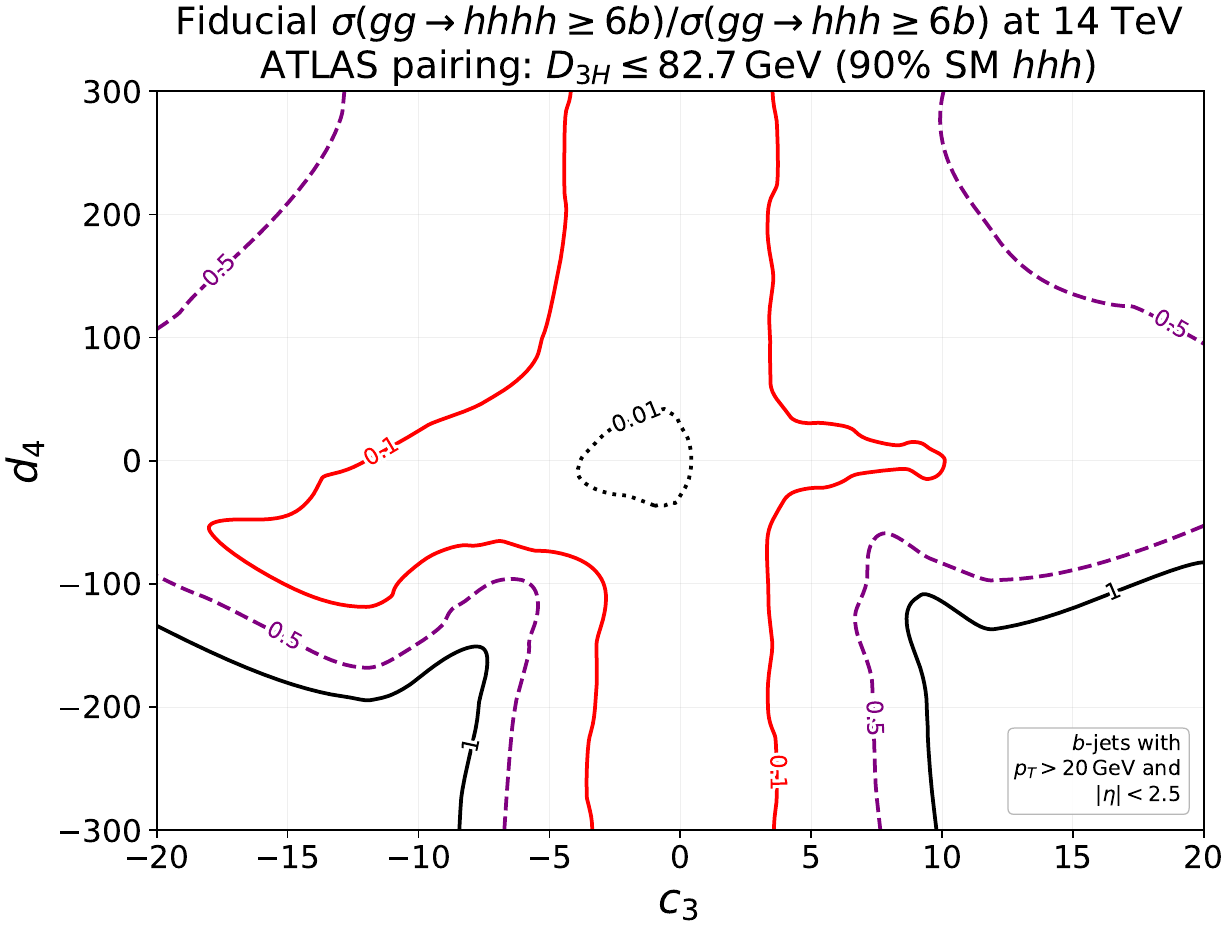}
  \caption{Contours of the paired fiducial cross-section ratio
    $R_{\rm fid}=\sigma_{\rm fid}(gg\to hhhh)/
    [\sigma_{\rm fid}(gg\to hhh)+\sigma_{\rm fid}(gg\to hhh+b\bar b)]$
    at 14~TeV on the $(c_3,d_4)$ plane. Both the numerator and denominator are
    selected in the inclusive ${\geq}6b$ category, requiring tagged $b$-jets
    with $p_T>20$~GeV and $|\eta|<2.5$, and must satisfy the loose pairing
    requirement $D_{3H}\leq82.7$~GeV. The pairing prescription follows the
    ATLAS reconstruction, while the threshold is defined in this study to
    retain $90\%$ of SM $hhh$ events passing the basic fiducial selection. The
    rates employ the same CMS-like jet-energy smearing and an analytic $85\%$
    per-jet tagging efficiency without mistags, and include the corresponding
    $h\to b\bar b$ branching fractions. The denominator combines the inclusive
    $hhh$ sample with the additive, unmatched $hhh+b\bar b$ estimate described
    in the text. The contours correspond to $R_{\rm fid}=0.01$, $0.1$, $0.5$,
    and $1$.}
  \label{fig:hhhh-hhh-fiducial-ratio}
\end{figure}

We further apply an ATLAS-inspired Higgs boson reconstruction. In each event, the six
highest-$p_T$ tagged jets are divided into three pairs, ordered by decreasing
pair $p_T$, by minimizing~\cite{ATLAS:2024xcs}
\begin{equation}
\begin{aligned}
D_{3H}=\min_{\rm pairings}\bigl(&|m_{bb,1}-120~{\rm GeV}|\\
 &+|m_{bb,2}-115~{\rm GeV}|\\
 &+|m_{bb,3}-110~{\rm GeV}|\bigr)\;.
\end{aligned}
\end{equation}
ATLAS employs this minimization to define the Higgs boson candidates and uses
their kinematics in a deep neural network (DNN) discriminant. Here, we impose a cut on $D_{3H}$ chosen to retain $90\%$ of the signal as a loose working point
for the purpose of the present estimate. Calibrating it on the SM $hhh$ sample after the
fiducial ${\geq}6b$ selection gives
$D_{3H}\leq82.7$~GeV and retains $90.0\%$ of the $gg \rightarrow hhh \rightarrow 6b$ events. We apply this fixed
threshold to all three processes and throughout the $(c_3,d_4)$ plane. Contours of the resulting ratio, $R_{\mathrm{fid}}$, are shown in
Fig.~\ref{fig:hhhh-hhh-fiducial-ratio}. We find that, in portions of the
sampled $(c_3,d_4)$ plane, the
fiducial $gg\to hhhh\to8b$ contribution can reach $10\%$--$100\%$ of the
combined $hhh$ and $hhh+b\bar b$ rate under the loose selection considered
here. This finding motivates a combined treatment of $hhh$ and $hhhh$ production
in future analyses of the inclusive ${\geq}6b$ final state, and could provide an initial avenue for sensitivity to $hhhh$
production even in the absence of a dedicated analysis.

\subsection{Extended Scalar Sector Analysis}
\label{sec:extended-scalar-analysis}

\subsubsection{AK4+AK8 Higgs Boson Reconstruction}

As the scalar-resonance masses increase, the Higgs bosons become more
energetic, and the two \(b\)-quarks from the decay of the same Higgs boson
become increasingly collimated. A reconstruction based only on eight resolved
jets therefore loses acceptance in the region of the mass scan where the Higgs bosons are highly boosted. We
instead reconstruct each event twice: once using anti-\(k_T\), \(R=0.4\) (AK4) jets
and then using anti-\(k_T\), \(R=0.8\) (AK8) jets~\cite{Cacciari:2008gp}, as introduced
in Sec.~\ref{sec:jet-smearing}. When the two \(b\)-quarks from a given
\(h\to b\bar b\) decay are sufficiently separated, they are reconstructed as
two distinct AK4 jets and paired to form a Higgs boson candidate. For a
sufficiently boosted Higgs boson, the \(b\bar b\) pair becomes collimated and
can instead be reconstructed within a single AK8 jet. Thus, an AK8 Higgs
candidate represents the same physical decay as a pair of resolved AK4 jets.
After the smearing described in Sec.~\ref{sec:jet-smearing}, AK4 jets are
required to satisfy \(p_T>20~\mathrm{GeV}\) and AK8 jets
\(p_T>300~\mathrm{GeV}\), with \(|\eta|<2.5\) in both cases. The AK8 jets are
groomed with the Soft Drop algorithm~\cite{Larkoski:2014wba}, using
\((\beta,z_{\rm cut})=(0,0.1)\). We impose no fixed groomed-mass window; the
groomed mass and the \(N\)-subjettiness ratio
\(\tau_{21}\)~\cite{Thaler:2010tr} are instead used by the multivariate
analysis.

A tagged AK8 jet provides a merged Higgs boson candidate, while pairs of
tagged AK4 jets provide resolved candidates. Whenever an AK8 candidate is
selected, AK4 jets within $\Delta R<0.8$ of it are removed to avoid counting
the same particles twice. From the remaining jets, we select the leading AK4
jets required to form exactly four Higgs boson candidates. For each tagging
outcome, we examine the possible AK4 pairings and order the four candidates in
each complete assignment by transverse momentum. As in the self-coupling
analysis, we retain the assignment that minimizes
$\sum_i(m_i-m_i^{\rm target})^2$, with
$(m_1^{\rm target},m_2^{\rm target},m_3^{\rm target},m_4^{\rm target})=
(120,115,110,105)~\mathrm{GeV}$. For an AK8 candidate this comparison uses
the Soft Drop mass, while its ungroomed four-vector defines the candidate
kinematics.

As in the self-coupling analysis, the tagging response is applied analytically
rather than through per-event random tag decisions. For AK4 jets we use $b$-, $c$- and
light-jet tagging probabilities of $\epsilon_b=0.85$, $\epsilon_c=0.10$ and
$\epsilon_j=0.01$, respectively, as previously. For AK8 jets we consider a
double-$b$ tag with efficiency $\epsilon_{bb}=\epsilon_b^2=0.7225$ for a
genuine $h\to b\bar b$ jet and a false-tag probability $f_{bb}=0.10$ for all
other AK8 jets.\footnote{This simplified parametrization neglects the
dependence of the tagging response on jet flavour and kinematics.
Nevertheless, the CMS study~\cite{CMS:2020ttz} reported that an AK8
double-$b$ tagging efficiency of approximately $75\%$ for simulated
$h\to b\bar b$ jets could be obtained with a mistag probability of
approximately $10\%$ in inclusive multijet events. The corresponding
probability for simulated $h\to c\bar c$ jets was approximately $33\%$,
illustrating the flavour dependence that is not modelled here. We therefore
treat $(\epsilon_{bb},f_{bb})=(0.7225,0.10)$ as an illustrative working point
rather than a detector-specific calibration.}

For each simulated event, we consider whether each of up to four leading AK8
jets passes or fails the double-$b$ tag. For example, if two such AK8 jets are
present, there are four possibilities: neither jet is tagged, only the first
is tagged, only the second is tagged, or both are tagged. In each case, the
tagged AK8 jets form merged Higgs boson candidates, the overlapping AK4 jets
are removed, and the remaining candidates are reconstructed from AK4-jet
pairs. The different cases can therefore produce different reconstructed
kinematics and classifier scores.

Each case is weighted by the probability of its AK8 tagging pattern and by the
relevant AK4 tagging or mistagging factors. We then sum these weighted
contributions. They represent the possible tagging results for one simulated
event, rather than several independent events.

\subsubsection{Fixed-Efficiency Event-Count Analysis}

The signal kinematics and reconstruction efficiency depend strongly on the
masses of the new scalars. We evaluate the resolved, mixed, and boosted
reconstruction at every generated mass point and add the three contributions.
We train one mass-conditioned \texttt{XGBoost} classifier for the direct topology,
\(pp\rightarrow S\rightarrow hhhh\), and one for the cascade topology,
\(pp\rightarrow h_3\rightarrow h_2h_2\rightarrow4h_1\), with $h_1\equiv h$.
All generated masses contribute to the relevant training, with equal total
signal weight assigned to every mass point. The tested value of $M_S$ or the
pair $(M_2,M_3)$ is supplied to the
classifier together with the reconstructed event information, following the
parametrized-classifier strategy of Ref.~\cite{Baldi:2016fzo}. The classifier can
therefore follow the transition from resolved AK4 reconstruction to mixed and
boosted AK4+AK8 reconstruction across the scans. The reconstructed observables
and the additional mass-hypothesis inputs are defined in~\ref{app:classifier-observables}.

Following the same five-group procedure as in the self-coupling analysis, we
assign each original simulated event to one of five fixed groups. All weighted
reconstructions obtained from the different AK8 tagging possibilities of the
same event are kept in the same group. In each of the five rotations, three
groups are used to train the classifier, one validation group determines the
score threshold, and the remaining group is kept aside to determine the
resonant signal efficiency. 

The actual use of the classifier score differs from that in the self-coupling
analysis. There, all five background-percentile intervals enter the likelihood
as separate event counts. Here, we instead retain a single event count above a
classifier-score threshold. We consider working points that retain $50\%$,
$30\%$ or $20\%$ of the validation background. At each mass point, we calculate
the expected cross-section upper limit for all three working points and compare
each result with the best one at that point. We then use the median relative
performance over the complete mass scan to select one background efficiency
for the direct topology and one for the cascade topology. If retaining more events worsens the median expected upper-limit sensitivity
by no more than $2\%$, we choose this looser working point. Otherwise, we use
the working point with the best expected sensitivity.

This procedure selects an optimal background efficiency of $20\%$ for both the direct
and cascade topologies, and we use this working point at every mass point. It
is the retained background fraction of $20\%$, rather than the numerical
\texttt{XGBoost}-score threshold, that remains fixed. For each mass hypothesis
and each rotation, the validation background determines the score threshold
above which its highest-scoring $20\%$ is retained. The threshold can therefore
differ between masses and rotations because the classifier and its score
distribution change. Once this threshold has been fixed, it is applied to the
held-out group to determine the resonant signal efficiency. The held-out
events are thus not used to choose the working point, and no mass point is
allowed to select a different background efficiency.

After completing the five rotations, we add the selected yields from the five
validation groups for the conventional backgrounds and the three SM
multi-Higgs boson processes. This gives the SM Asimov event count at each mass
hypothesis. We similarly add the selected resonant-signal yields from the five
held-out groups to determine the expected signal yield for a cross section of
$1~\mathrm{fb}$. Each original event, together with all its weighted
reconstructions, contributes through only one group to the corresponding
total.

The resolved, mixed, and boosted contributions are not treated as separate
likelihood categories. Instead, their selected yields are combined into one
signal-enriched event count, from which we calculate the expected
cross-section upper limit. This preserves the mass-dependent discrimination
provided by the classifier without dividing the available weighted background
sample into several sparsely populated score intervals. The use of a single selected event count is therefore a practical compromise
imposed by the available Monte Carlo background statistics. Generating
sufficiently large samples of the high-multiplicity backgrounds to populate
several classifier-score intervals reliably at every mass hypothesis is
computationally demanding. With larger background samples, a multi-bin fit
could use more of the classifier information and provide additional
sensitivity.

Comparing our strategy with the resonant ATLAS $hhh\to6b$ analysis, the main
similarity is that both use a mass-dependent classifier to identify
signal-enriched events, rather than relying on an inclusive event count.
ATLAS searches for $X\to Sh\to hhh$, where $S$ and $X$ are new scalars, using
resolved AK4 jets and several classifier-score intervals. It also constrains
the backgrounds using data control regions and includes a full uncertainty
model~\cite{ATLAS:2024xcs}. In contrast, we study direct and cascade production
of four Higgs bosons, combine AK4 and AK8 reconstructions, and use one selected
score region in a statistics-only Poisson comparison. 

Physical event rates are calculated independently of the classifier training
weights. We use an integrated luminosity of $\mathcal{L} = 3000~\mathrm{fb}^{-1}$,
$\mathrm{BR}(h\rightarrow b\bar b)=0.5824$ and a common background $K$-factor
of two. The resonant signal template is normalized to a production rate of
1~fb into four Higgs bosons before their decays, with no additional signal
$K$-factor. The two signal rates are therefore defined as
\begin{equation}
\begin{aligned}
  \sigma_{\mathrm{dir}} &\equiv \sigma(pp\rightarrow S) \times 
  \mathrm{BR}(S\rightarrow 4h), \\
  \sigma_{\mathrm{cas}} &\equiv \sigma(pp\rightarrow h_3) \times
  \mathrm{BR}(h_3\rightarrow h_2h_2) \\
  &\quad\times
  \mathrm{BR}(h_2\rightarrow h_1h_1)^2.
\end{aligned}
\label{eq:resonance-rate-definitions}
\end{equation}
Both definitions precede the decays of the four SM-like Higgs bosons to $b\bar b$.

In the present extended scalar sector scenario, the SM multi-Higgs boson processes $H_{\rm sel}^{hhh b\bar b}(m)$ and $H_{\rm sel}^{hh+4b}(m)$ are part of the expected event count, i.e.\ they are considered as backgrounds rather than signal. At mass hypothesis $m$, the selected
Asimov count is
\begin{equation}
\begin{aligned}
 n(m)={}&B_{\rm sel}(m)+H_{\rm sel}^{hhhh}(m)\\
 &+H_{\rm sel}^{hhh b\bar b}(m)\\
 &+H_{\rm sel}^{hh+4b}(m)\;,
\end{aligned}
 \label{eq:resonance-asimov-count}
\end{equation}

where $B_{\rm sel}$ collects all the backgrounds that do not originate from multi-Higgs boson (two or more) processes. 

If $S_{\rm sel}^{1\,\mathrm{fb}}(m)$ is the selected resonant yield, normalized to
1~fb before the four Higgs bosons decay, the tested count is
\begin{equation}
 \begin{aligned}
 \nu(\sigma;m)&=n(m)+
 \frac{\sigma}{1~\mathrm{fb}}S_{\rm sel}^{1\,\mathrm{fb}}(m)\;,\\
 m&=M_S\ \hbox{or}\ (M_2,M_3)\;.
 \end{aligned}
 \label{eq:resonance-tested-count}
\end{equation}
We use the same Poisson comparison of predicted event counts as in
Eq.~\eqref{eq:self-coupling-poisson-q},
\begin{equation}
 q(\sigma;m)=2\left[
 \nu(\sigma;m)-n(m)+n(m)\ln\frac{n(m)}{\nu(\sigma;m)}
 \right]\;,
 \label{eq:resonance-poisson-q}
\end{equation}
and solve $q(\sigma_{95};m)=3.841$.\footnote{Although the upper limit is
one-sided, the value $q=3.841$ follows from the asymptotic
$\mathrm{CL}_s$ construction rather than from the unmodified
$p_{\sigma}$ criterion. For a background-only Asimov data set,
$\mathrm{CL}_s=2[1-\Phi(\sqrt{q})]$, where $\Phi$ denotes the
standard-normal cumulative distribution. Hence,
$\mathrm{CL}_s=0.05$ gives
$q=[\Phi^{-1}(0.975)]^2=3.841$~\cite{Cowan:2010js,Read:2002hq}.} Fractional expected event counts are used
directly. The background and SM multi-Higgs boson rates are treated as known; no
nuisance parameters are introduced. These are therefore statistics-only
expected limits.

The direct topology is presented as $\sigma_{95}(M_S)$ and the cascade
topology as $\sigma_{95}(M_2,M_3)$ over the sampled physical region. These are
limits on the production rates in Eq.~\eqref{eq:resonance-rate-definitions};
they become mass exclusions only after comparison with a specified model
prediction.

The generated samples permit an additional, more restricted
interpretation in terms of the interactions displayed in the simplified
Lagrangians of Eqs.~\eqref{eq:Ldir} and~\eqref{eq:Lcas} for the direct and cascade topologies, respectively. The widths of $S$, $h_2$ and $h_3$ are fixed externally to
$1~\mathrm{GeV}$ at every mass point and are not calculated from the
couplings. The largest width-to-mass ratio in either scan is
$3.6\times10^{-3}$, so these samples describe narrow resonances. Moreover,
all scalar interactions other than those required for the displayed direct
or cascade processes, or for the SM multi-Higgs boson processes, are set to zero. The corresponding cross sections of Eq.~\eqref{eq:resonance-rate-definitions} therefore
contain one common coupling combination:
\begin{equation}
\begin{aligned}
\sigma_\mathrm{dir}(M_S) &= \tilde{\sigma}_\mathrm{dir}(M_S) \times \left(\frac{C_{ggS}c_5}{\Lambda^2}\right)^2  \;, \\
\sigma_\mathrm{cas}(M_2, M_3) &= \tilde{\sigma}_\mathrm{cas}(M_2, M_3) \times  \left(\frac{C_{gg3}\lambda_{223}\lambda_{112}^{2}}{\Lambda} \right)^2\;, \\
\end{aligned}
\end{equation}
where the $\tilde{\sigma}$ are the reduced LO cross-section coefficients at
fixed widths and for the fixed collider and generator setup. When
$\rho_{\rm dir}$ and $\rho_{\rm cas}$ are expressed in $\mathrm{TeV}^{-2}$
and $\mathrm{TeV}^{2}$, respectively, $\tilde{\sigma}_{\rm dir}$ has units of
$\mathrm{fb}\,\mathrm{TeV}^{4}$ and $\tilde{\sigma}_{\rm cas}$ has units of
$\mathrm{fb}\,\mathrm{TeV}^{-4}$. We retain the native mass dimensions of
the coupling combinations and define
\begin{equation}
\begin{aligned}
 \rho_{\rm dir} &\equiv
 \frac{|C_{ggS}c_5|}{\Lambda^2},\\
 \rho_{\rm cas} &\equiv
 \frac{|C_{gg3}\lambda_{223}\lambda_{112}^{2}|}{\Lambda}.
\end{aligned}
\label{eq:resonance-coupling-products}
\end{equation}
Thus, $\rho_{\rm dir}$ contains the effective gluon--$S$ coefficient
$C_{ggS}$, the coefficient $c_5$ of the $Sh^4$ interaction and the effective
scale $\Lambda$. The cascade combination $\rho_{\rm cas}$ contains the
gluon--$h_3$ coefficient $C_{gg3}$, the two scalar couplings $\lambda_{223}$
and $\lambda_{112}$, and $\Lambda$. Thus, $\rho_{\rm dir}$ has mass dimension
$-2$, while $\rho_{\rm cas}$ has mass dimension $+2$; below we quote them in
$\mathrm{TeV}^{-2}$ and $\mathrm{TeV}^{2}$, respectively. At fixed masses and
widths, the signal cross section is proportional to the square of the
corresponding combination. Hence, for either topology $X$,
\begin{equation}
 \rho_{95,X}(m)=
 \left[\frac{\sigma_{95,X}(m)}
 {\tilde{\sigma}_X(m)}\right]^{1/2},
 \qquad X\in\{\mathrm{dir},\mathrm{cas}\},
 \label{eq:resonance-coupling-rescaling}
\end{equation}
The reduced coefficients $\tilde{\sigma}_X(m)$ are obtained point by point
from the corresponding LHE cross sections after factoring out the squared
coupling combination evaluated with the event-generation-card parameters.
These parameters set only the matrix-element normalization; they are not
treated as a physical benchmark. The universal QCD factor in the gluon
interaction is already included in $\tilde{\sigma}_X$.
Eq.~\eqref{eq:resonance-coupling-rescaling} therefore uses the same event
selection and statistical limit without a new classifier or detector
simulation.

This conversion is conditional on the fixed $1~\mathrm{GeV}$ widths, the LO
signal normalization and the absence of additional diagrams and interference.
If the same couplings determine the resonance widths, the branching fractions
in Eq.~\eqref{eq:resonance-rate-definitions} must instead be calculated in the
specified model. The rate also constrains only the magnitude of each combination,
not its sign. We consequently retain the cross-section limits as the primary
results and present the coupling combinations below as an illustrative
interpretation.

\subsubsection{Expected Resonance Cross-Section Limits}
\label{sec:extended-scalar-results}

All 42 direct and 441 cascade mass points produce finite, positive limits. Table~\ref{tab:resonance-score-yields} shows the input and classifier-selected
rates at representative direct and cascade mass hypotheses. The input columns
make the sample normalization explicit, while the output columns demonstrate
the effect of the mass-conditioned \texttt{XGBoost} selection. These input rates are
not directly comparable to the resolved-only rates in
Table~\ref{tab:self-coupling-score-yields}: here the hybrid jet reconstruction
also accepts mixed and boosted configurations containing double-tagged AK8
jets. The generator cross sections themselves are common to the two analyses.

\begin{table*}[t]
\centering
\footnotesize
\setlength{\tabcolsep}{3.0pt}
\renewcommand{\arraystretch}{1.06}
\begin{tabularx}{\textwidth}{@{}Yrrrrrr@{}}
\toprule
\textbf{Sample} &
\shortstack{\boldmath $\sigma_{\rm input}^{\mathrm{AK4+AK8}}$\\$[\mathrm{fb}]$} &
\shortstack{\boldmath $N_{\rm input}^{\mathrm{AK4+AK8}}$\\\textbf{expected}} &
\shortstack{\boldmath $\sigma_{\rm XGB}^{\rm dir}$\\$[\mathrm{fb}]$} &
\shortstack{\boldmath $N_{\rm XGB}^{\rm dir}$\\\textbf{expected}} &
\shortstack{\boldmath $\sigma_{\rm XGB}^{\rm cas}$\\$[\mathrm{fb}]$} &
\shortstack{\boldmath $N_{\rm XGB}^{\rm cas}$\\\textbf{expected}} \\
\midrule
\multicolumn{7}{@{}l}{\emph{Representative resonant signal points}} \\
Direct resonant signal, $M_S=1.5~\mathrm{TeV}$ & $6.82\!\times\!10^{-3}$ & 20.45 & $6.46\!\times\!10^{-3}$ & 19.39 & -- & -- \\
Cascade resonant signal, $(M_2,M_3)=(625,1500)~\mathrm{GeV}$ & $8.18\!\times\!10^{-3}$ & 24.54 & -- & -- & $7.43\!\times\!10^{-3}$ & 22.29 \\
\midrule
\multicolumn{7}{@{}l}{\emph{Standard Model multi-Higgs-boson and background samples}} \\
SM $gg\to hhhh$ & $8.13\!\times\!10^{-7}$ & $2.44\!\times\!10^{-3}$ & $4.61\!\times\!10^{-7}$ & $1.38\!\times\!10^{-3}$ & $3.71\!\times\!10^{-7}$ & $1.11\!\times\!10^{-3}$ \\
SM $gg\to hhh+b\bar b$ & $2.73\!\times\!10^{-6}$ & $8.19\!\times\!10^{-3}$ & $1.6\!\times\!10^{-6}$ & $4.79\!\times\!10^{-3}$ & $1.3\!\times\!10^{-6}$ & $3.91\!\times\!10^{-3}$ \\
SM $gg\to hh+4b$ & $3.94\!\times\!10^{-4}$ & 1.182 & $3.37\!\times\!10^{-4}$ & 1.011 & $3.21\!\times\!10^{-4}$ & 0.963 \\
\textbf{Total SM multi-Higgs-boson} & \textbf{$3.98\!\times\!10^{-4}$} & \textbf{1.193} & \textbf{$3.39\!\times\!10^{-4}$} & \textbf{1.017} & \textbf{$3.23\!\times\!10^{-4}$} & \textbf{0.9681} \\
\addlinespace
$gg\to8b$ & $5.48\!\times\!10^{-3}$ & 16.45 & $2.43\!\times\!10^{-3}$ & 7.291 & $1.92\!\times\!10^{-3}$ & 5.768 \\
$gg\to6b+2j$ & 0.06208 & 186.2 & 0.04503 & 135.1 & 0.0403 & 120.9 \\
$gg\to6b+c\bar c$ & $7.55\!\times\!10^{-4}$ & 2.264 & $4.19\!\times\!10^{-4}$ & 1.257 & $3.66\!\times\!10^{-4}$ & 1.097 \\
$pp\to Z+6b$, $Z\to b\bar b$ & $3.88\!\times\!10^{-4}$ & 1.165 & $2.03\!\times\!10^{-4}$ & 0.6091 & $1.67\!\times\!10^{-4}$ & 0.4996 \\
$gg\to4b+4j$ & 1.335 & $4\!\times\!10^{3}$ & 0.1362 & 408.6 & 0.2117 & 635 \\
$gg\to4b+c\bar c+2j$ & 0.03229 & 96.88 & $5.24\!\times\!10^{-3}$ & 15.72 & 0.02432 & 72.95 \\
$gg\to4b+4c$ & $8.31\!\times\!10^{-5}$ & 0.2494 & $2.63\!\times\!10^{-5}$ & 0.07879 & $2.1\!\times\!10^{-5}$ & 0.06291 \\
$gg\to h+6b$, $h\to b\bar b$ & $3.2\!\times\!10^{-6}$ & $9.6\!\times\!10^{-3}$ & $2.03\!\times\!10^{-6}$ & $6.09\!\times\!10^{-3}$ & $1.76\!\times\!10^{-6}$ & $5.29\!\times\!10^{-3}$ \\
$t\bar t+4b+2c+2j$ & $2.01\!\times\!10^{-4}$ & 0.6035 & $1.32\!\times\!10^{-4}$ & 0.3951 & $1.14\!\times\!10^{-4}$ & 0.341 \\
$t\bar t+4b+c+3j$ & $3.15\!\times\!10^{-4}$ & 0.9449 & $2.1\!\times\!10^{-4}$ & 0.6315 & $1.81\!\times\!10^{-4}$ & 0.5419 \\
$t\bar t+4b+4j$ & $1.19\!\times\!10^{-4}$ & 0.3582 & $8.76\!\times\!10^{-5}$ & 0.2627 & $7.25\!\times\!10^{-5}$ & 0.2174 \\
\textbf{Total conventional background} & \textbf{1.437} & \textbf{$4.31\!\times\!10^{3}$} & \textbf{0.19} & \textbf{570} & \textbf{0.2791} & \textbf{837.4} \\
\midrule
\textbf{Asimov total} & \textbf{1.437} & \textbf{$4.31\!\times\!10^{3}$} & \textbf{0.1903} & \textbf{571} & \textbf{0.2795} & \textbf{838.4} \\
\bottomrule
\end{tabularx}
\caption{Input cross sections and event counts, together with the direct and cascade \texttt{XGBoost} outputs, at $\sqrt{s}=14~\mathrm{TeV}$ and $\mathcal{L}=3000~\mathrm{fb}^{-1}$. The input columns give the common hybrid tagged AK4+AK8 preselection after the object requirements, branching fractions, $K$-factors and analytic tagging or mistagging probabilities, but before Higgs boson candidate pairing and \texttt{XGBoost}. Thus, $\sigma_{\rm input}^{\mathrm{AK4+AK8}}$ is a post-tagging effective cross section, not the hard-process generator cross section, and $N_{\rm input}^{\mathrm{AK4+AK8}}=\mathcal{L}\sigma_{\rm input}^{\mathrm{AK4+AK8}}$. A genuine double-$b$ AK8 jet is weighted by $\epsilon_{bb}=\epsilon_b^2=0.7225$, a false double tag by $f_{bb}=0.10$, and each double-tagged AK8 candidate replaces two required AK4 tags. The underlying generator cross sections and $K$-factors are common to the two
analyses. The different input rates arise from the hybrid AK4+AK8 selection
used here, rather than the resolved-only selection used in
Table~\ref{tab:self-coupling-score-yields}. The direct and cascade \texttt{XGBoost} columns use the $20\%$ target background-efficiency working point and the mass hypothesis of the corresponding reference signal. They give the cross section and expected yield in the single selected classifier-score region after the complete resolved, mixed and boosted reconstruction. The resonant inputs assume a production rate of 1~fb before the four Higgs boson decays. The direct and cascade \texttt{XGBoost} columns show the yields selected by the corresponding classifiers. Each resonant signal is evaluated only for its own
topology, with dashes marking combinations that are not considered. The Asimov
total includes the conventional backgrounds and the three Standard Model
multi-Higgs boson processes. Totals may differ slightly from the sums of the
displayed entries because of rounding.}
\label{tab:resonance-score-yields}
\end{table*}

Separately from the total rates presented in
Table~\ref{tab:resonance-score-yields}, we examine how the reconstructed and
tagged signal yield is divided among the resolved, mixed, and boosted
categories before the classifier selection. These categories are combined in
the event count used for the limit, but their relative contributions illustrate
the transition from AK4- to AK8-dominated reconstruction. For direct
production, the resolved category accounts for $95\%$ of the signal yield at
$M_S=600$~GeV, while the mixed category contributes $68\%$ at
$M_S=1.5$~TeV and the boosted category contributes $74\%$ at $M_S=4$~TeV.
For the cascade benchmarks, the dominant contributions are $97\%$ resolved at
$(M_2,M_3)=(275,600)$~GeV, $66\%$ mixed at $(625,1500)$~GeV, and $72\%$
boosted at $(1500,3500)$~GeV.

\begin{figure*}[t]
\centering
\includegraphics[width=\textwidth]{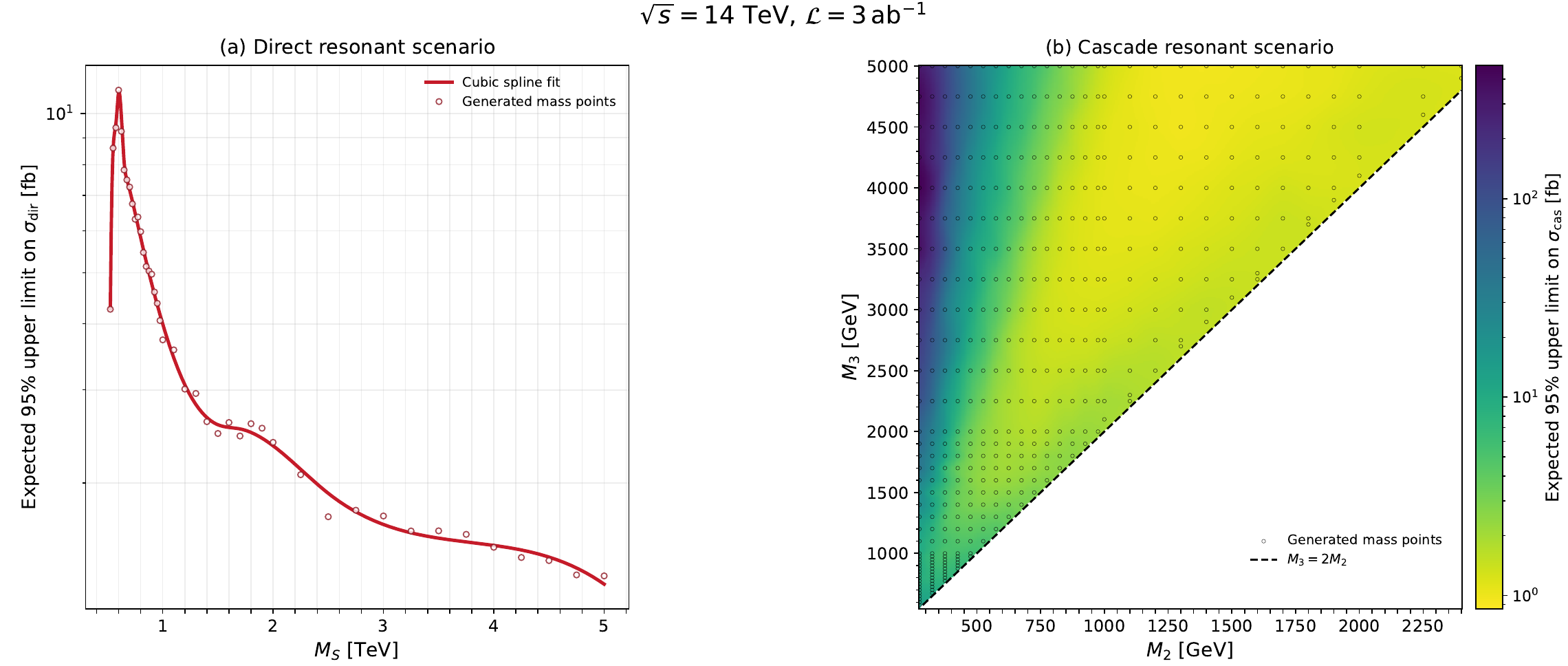}
\caption{Expected fixed-background-efficiency one-bin 95\% upper limits on the
resonant quadruple-Higgs boson production rates. Panel~(a) shows
$\sigma_{\rm dir}$ as a function of $M_S$; every marker is a generated mass
point and the red curve is a cubic
smoothing-spline fit to the logarithms of the pointwise limits, included only
as a visual guide. Panel~(b) shows
$\sigma_{\rm cas}$ over the $(M_2,M_3)$ plane.
Open circles mark the generated mass points and the dashed line shows the
kinematic boundary $M_3=2M_2$. The conservative colour surface is used only
for visualization, is continued to the kinematic boundary, and is not extended
into the unphysical region.}
\label{fig:resonance-limits}
\end{figure*}

\paragraph{\boldmath $pp\rightarrow S\rightarrow4h$.}
Panel~(a) of Fig.~\ref{fig:resonance-limits} shows the expected upper limit at
each of the 42 generated masses. At $M_S=1.5$~TeV, the limit is
$\sigma_{\rm dir}<2.48$~fb at the 95\% confidence level. Across the scan it ranges
from 1.33~fb at $M_S=5$~TeV to 11.1~fb at $M_S=600$~GeV. The markers show the
exact fixed-efficiency one-bin limits at the physical mass points. The red curve is a cubic
smoothing-spline fit to their logarithms and is included only as a visual
guide; it is not used to calculate or quote a limit at any mass.

The turn at the lowest masses is driven mainly by the signal reconstruction and
tagging acceptance. For a $1~\mathrm{fb}$ signal, the selected yield decreases
from 12.04 events at $M_S=525$~GeV to 4.48 events at $600$~GeV, before
increasing again at larger masses. Since the classifier retains at least
$93.7\%$ of the reconstructed signal in this region, this behaviour arises
mainly before the score selection and reflects the rapidly changing event
kinematics near the quadruple-Higgs boson production threshold.

\paragraph{\boldmath $pp\rightarrow h_3\rightarrow h_2h_2\rightarrow4h_1$.}
At the reference point $(M_2,M_3)=(625,1500)$~GeV, the expected limit is
$\sigma_{\rm cas}<2.60$~fb. Across the generated plane, the strongest
pointwise limit is 0.857~fb at $(1400,5000)$~GeV, while the weakest is
469~fb at $(275,4000)$~GeV, where the
four separate Higgs boson candidates are difficult to retain. The complete
result is shown in panel~(b) of Fig.~\ref{fig:resonance-limits}.

For visualization, the colour map uses a smooth, conservative interpolation of
$\log_{10}\sigma_{95}$. At each generated mass point, it never shows a smaller
upper limit than the calculated value. The
open circles show the 441 points used for the numerical results; the
interpolation is applied only to the displayed limit surface, not to the signal
or background yields.

\begin{figure*}[t]
\centering
\includegraphics[width=\textwidth]{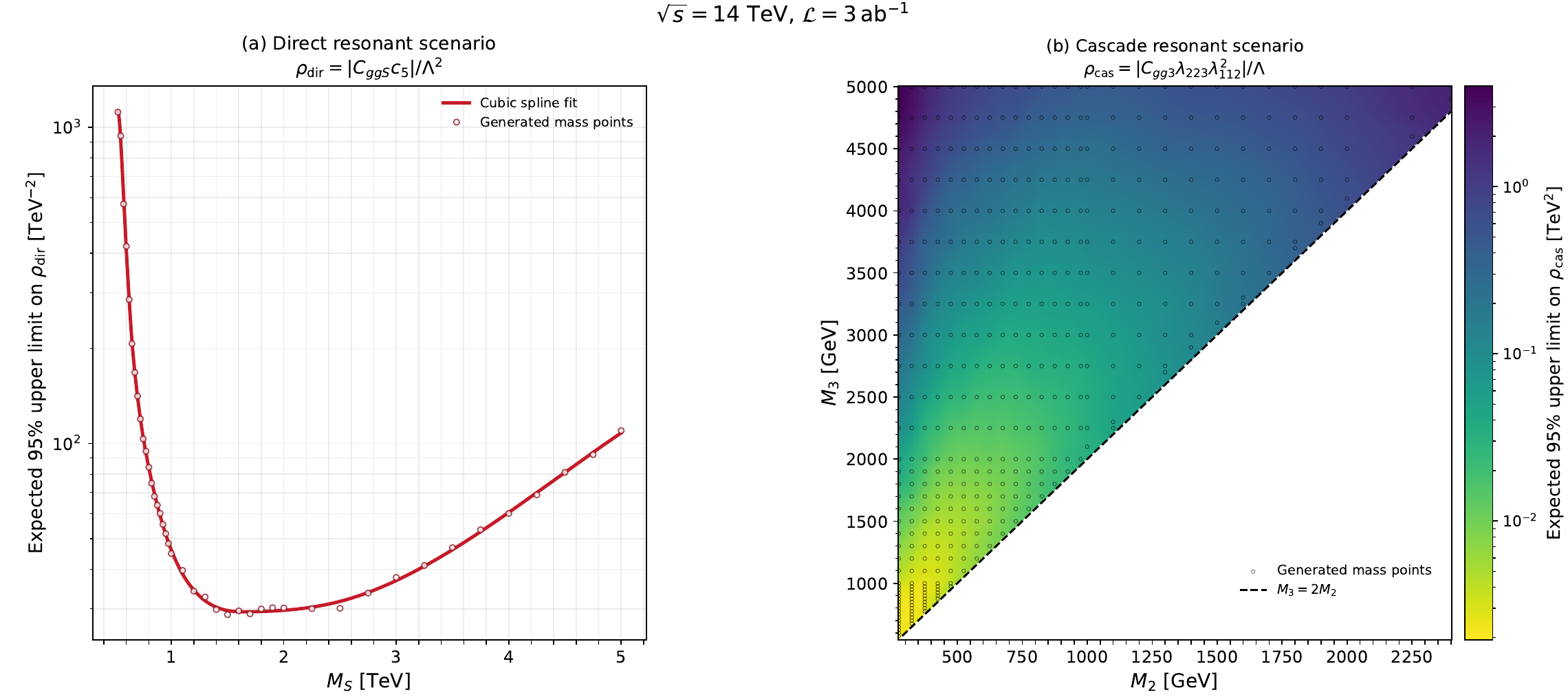}
\caption{Illustrative expected 95\% upper limits on the dimensionful coupling
combinations $\rho_{\rm dir}$ and $\rho_{\rm cas}$ defined in
Eq.~\eqref{eq:resonance-coupling-products}. The direct and cascade results are
shown in $\mathrm{TeV}^{-2}$ and $\mathrm{TeV}^{2}$, respectively, and the
coupling variables entering each combination are displayed above the
corresponding panel. Panel~(a)
assumes $\Gamma_S=1~\mathrm{GeV}$; every marker is obtained from the exact
pointwise cross-section limit and the reduced cross-section coefficient
extracted from the LHE sample at the same mass, while the red curve is a cubic
smoothing-spline fit to the logarithms of these values and is included only as
a visual guide. Panel~(b) assumes
$\Gamma_{h_2}=\Gamma_{h_3}=1~\mathrm{GeV}$; open circles mark the exact
generated mass points and the dashed line shows the boundary $M_3=2M_2$. The
conservative colour surface is used only for visualization and is continued to
the kinematic boundary.}
\label{fig:resonance-rho-limits}
\end{figure*}

\paragraph{Illustrative fixed-width coupling combinations.}
Applying Eq.~\eqref{eq:resonance-coupling-rescaling} point by point gives the
direct result in panel~(a) of Fig.~\ref{fig:resonance-rho-limits}. At
$M_S=1.5~\mathrm{TeV}$, we obtain the strongest bound in the scan,
$\rho_{\rm dir}<28.74~\mathrm{TeV}^{-2}$ at the 95\% confidence level. The bound
weakens to $1.12\times10^3~\mathrm{TeV}^{-2}$ at the lowest generated mass,
$M_S=525~\mathrm{GeV}$. Its mass dependence is not determined by the
cross-section limit alone, but also by the reduced coefficient
$\tilde{\sigma}_{\rm dir}(M_S)$ entering
Eq.~\eqref{eq:resonance-coupling-rescaling}. This coefficient is suppressed by
the four-body phase space near the threshold for producing four Higgs bosons
and by the falling gluon luminosity at high masses, weakening the coupling
bound at both ends of the scan.

For the cascade benchmark with $M_2=625~\mathrm{GeV}$ and
$M_3=1.5~\mathrm{TeV}$, the corresponding result is
$\rho_{\rm cas}<6.76\times10^{-3}~\mathrm{TeV}^{2}$. Across the
plane, the strongest bound is
$1.94\times10^{-3}~\mathrm{TeV}^{2}$ at $(325,750)~\mathrm{GeV}$, and the
weakest bound is $4.00~\mathrm{TeV}^{2}$ at $(275,5000)~\mathrm{GeV}$. The full
result is shown in panel~(b) of Fig.~\ref{fig:resonance-rho-limits}.
In panel~(a), the markers retain the exact values of
$\rho_{\rm dir}$, while the red curve is a cubic smoothing-spline fit to their
logarithms. As in Fig.~\ref{fig:resonance-limits}, the fit is used only as a
visual guide and does not enter the quoted limits.

\FloatBarrier
\section{Conclusions}\label{sec:conclusions}

We have investigated quadruple-Higgs boson production at the high-luminosity LHC,
at $\sqrt{s}=14$~TeV and with $\mathcal{L}=3000~\mathrm{fb}^{-1}$, using the
eight-$b$-jet final state. We have considered both its sensitivity to the
trilinear and quartic Higgs boson self-couplings and its application to
extended scalar sectors containing resonances that decay into four Higgs
bosons, either directly or via cascade decays.

For the self-coupling analysis, we combine the $gg\to hhhh$,
$gg\to hhh+b\bar b$, and $gg\to hh+4b$ contributions, retaining their
different dependences on $c_3$ and $d_4$. The multivariate classifier output is divided
into five regions, all of which enter the event-count comparison. The expected
simultaneous 95\% confidence region projects onto
\[
-10.3<c_3<11.2,\qquad -222<d_4<234.
\]
For context, the projection of the no-systematics simultaneous ATLAS
$hhh\to6b$ contour of Ref.~\cite{ATLAS:2025cae} has a comparable reach in
$c_3$ and a narrower range in $d_4$. When one coupling is fixed to its SM
value, we obtain $-191<d_4<197$ for $c_3=0$ and $-7.6<c_3<10.3$ for $d_4=0$.
For the corresponding one-parameter comparison at $c_3=0$, ATLAS quotes
$-80<d_4<89$, which is narrower than our interval by roughly a factor of two.
The two results are nevertheless complementary, since they use different
signal-region definitions and different combinations of multi-Higgs boson
processes.

At the SM point, the resolved eight-tag selection contains $0.394$ expected
signal events and $21.9$ background events. Approximately $98\%$ of this
signal is supplied by $hh+4b$, rather than by $hhhh$ itself. For sizeable
departures from the SM self-couplings, the $hhhh$ and $hhh+b\bar b$
components become more important and provide much of the sensitivity. The
selected final state should therefore be regarded as a high-multiplicity
multi-Higgs boson region, rather than as an exclusive measurement of
quadruple-Higgs boson production.

The same observation is relevant to inclusive triple-Higgs boson searches.
Because the ATLAS $hhh\to6b$ signal region accepts events with at least six
tagged jets, $hhhh\to8b$ events can enter it directly. In our loose fiducial
study, the quadruple-Higgs boson contribution reaches between $10\%$ and $100\%$
of the combined $hhh$ and $hhh+b\bar b$ rate in parts of the $(c_3,d_4)$
plane. Including $hhhh$ as an explicit signal component in an inclusive
${\geq}6b$ analysis could therefore provide a practical first route to
quadruple-Higgs boson production without introducing a separate event
category.

For the extended scalar sectors, we considered two resonant production
mechanisms. The direct scenario is motivated by a one-real-singlet extension
and contains a new scalar $S$, produced through an effective coupling to
gluons and decaying as $pp\to S\to hhhh$. The cascade scenario is motivated
by a two-real-singlet extension and proceeds through
$pp\to h_3\to h_2h_2\to(h_1h_1)(h_1h_1)$, where $h_1$ is the SM-like Higgs
boson. We combine resolved AK4 and boosted AK8 reconstruction to cover the
changing event kinematics across the mass scans. For the direct benchmarks,
the reconstructed and tagged signal yield before the classifier selection
changes from $95\%$ resolved at $M_S=600$~GeV to $68\%$ mixed at
$1.5$~TeV and $74\%$ boosted at $4$~TeV, illustrating the role of the
combined AK4+AK8 reconstruction. The expected 95\% upper limits on the direct
production rate lie between $1.33$ and $11.1$~fb over the generated mass
range, with $\sigma_{\rm dir}<2.48$~fb at $M_S=1.5$~TeV. For the cascade
topology, the limits range from $0.857$ to $469$~fb across the $(M_2,M_3)$
plane, with $\sigma_{\rm cas}<2.60$~fb at
$(M_2,M_3)=(625,1500)$~GeV. These are cross-section limits rather than mass
exclusions; a mass exclusion requires a prediction for the production rate
and branching fractions in a specified model. The precise values of these cross-section limits for both scenarios are presented in ancillary files for convenience. 

For illustration, we have also translated the cross-section limits into the
dimensionful coupling combinations defined in
Eq.~\eqref{eq:resonance-coupling-products}. For resonance widths fixed
externally to $1$~GeV, the representative limits are
$\rho_{\rm dir}<28.74~\mathrm{TeV}^{-2}$ at $M_S=1.5$~TeV and
$\rho_{\rm cas}<6.76\times10^{-3}~\mathrm{TeV}^{2}$ at
$(M_2,M_3)=(625,1500)$~GeV. These values are conditional on the fixed-width
scenarios. Their interpretation in a complete extended scalar model requires
the widths and branching fractions to be calculated from the same couplings.
Such an interpretation must also account for additional diagrams, their
interference with non-resonant $gg\to hhhh$ production, and any additional
new-physics backgrounds. Mapping these limits onto specific
singlet extensions, in which the masses, widths, branching fractions, and
interference contributions are treated consistently, is a natural
continuation of this work.

The results presented here are statistics-only projections. Detector effects
are represented through jet-energy smearing and fixed tagging probabilities,
rather than a full detector simulation, and pile-up, trigger effects, and
systematic uncertainties are not included. The tagging probabilities should
therefore be regarded as illustrative working points rather than
detector-specific calibrations. The $B/4$ and $4B$ calculations show the
dependence of the self-coupling result on the assumed background level, but
do not constitute an uncertainty band. In addition, the $hh+4b$ score shape
and selection efficiency are evaluated from an SM sample in the
$m_t\to\infty$ limit, while its $c_3$-dependent normalization is supplied by
Eq.~\eqref{eq:hh4b-c3-fit}. The $hhh+b\bar b$ contribution uses a forced
parton-shower splitting of the full $hhh+g$ process.

A more complete analysis will require improved calculations of these
processes, larger simulated or data-constrained background samples, and an
experimental treatment of the background uncertainties. The large multijet
cross sections make the generation of sufficiently large background samples
a central practical limitation. With larger samples, the resonance analysis
could also use several classifier-score regions instead of one selected event
count. Further improvements in both triple- and quadruple-Higgs boson
production could come from reducing the combinatorial ambiguity in pairing
resolved $b$-jets into Higgs boson candidates. Identifying the $b$-jet charge
may provide one route; see, e.g., Ref.~\cite{CMS-DP-2025-071}. We leave this
to future work.

Quadruple-Higgs boson production is unlikely, by itself, to provide a precise
determination of the Higgs boson self-couplings at the high-luminosity LHC.
Nevertheless, it can make a non-negligible contribution to inclusive
high-multiplicity $b$-jet signal regions and provides sensitivity to scalar
resonances that decay into several Higgs bosons. Taken together, these results
motivate analyses in which all multi-Higgs boson processes contributing to a
given high-multiplicity $b$-jet signal region are modelled consistently.

\section*{Acknowledgements}
We would like to thank Liza Brost for useful discussions. AP acknowledges support by the National Science Foundation under Grant No.\ PHY 2210161 and the US Department of Energy, Office of Science, Office of Nuclear Physics under Award Number DE-SC0025728. G.T.-X. is supported by the European Union's Horizon 2020 research and
innovation programme under the Marie Skłodowska-Curie grant agreement
No.~945422. This research was supported by the Deutsche
Forschungsgemeinschaft (DFG, German Research Foundation) under grant
396021762 -- TRR~257.

\appendix
\renewcommand*{\theHequation}{\Alph{section}.\arabic{equation}}

\section{Approximate Simulation of Multi-Higgs Boson Non-Signal Processes}\label{app:gsplit}

Simulating the full loop-induced $gg\to hhh b\bar b$ process with the full
top-mass dependence is not computationally tractable for the present study.
We therefore employ an approximate description based on the full-loop
$gg\to hhh g$ matrix element, followed by a forced final-state
\(g\to b\bar b\) splitting in the \texttt{HERWIG} parton shower. The hard
process is generated with \texttt{MG5\_aMC}, retaining the full loop
dependence of the triple-Higgs boson production amplitude, while the additional
heavy-flavour pair is described by the shower splitting kernel.

We require the forced splitting to produce a $b\bar b$ pair satisfying
\(p_{T,b}>15~\mathrm{GeV}\), \(|\eta_b|<3.0\), and
\(\Delta R_{bb}>0.3\), with the same separation imposed relative to the other
final-state $b$-quarks, and we correct the event weights by the measured
forced-splitting acceptance. The Higgs bosons are subsequently decayed as
$h\to b\bar b$, so that this sample contributes to the same reconstructed
$8b$ final state as the $gg\to hhhh$ signal. This procedure does not reproduce
the complete $gg\to hhh b\bar b$ matrix element, in particular away from the
collinear region, and is treated as an approximate shower-based estimate.
\begin{figure*}[t]
  \centering
\includegraphics[height=0.22\textheight,keepaspectratio]{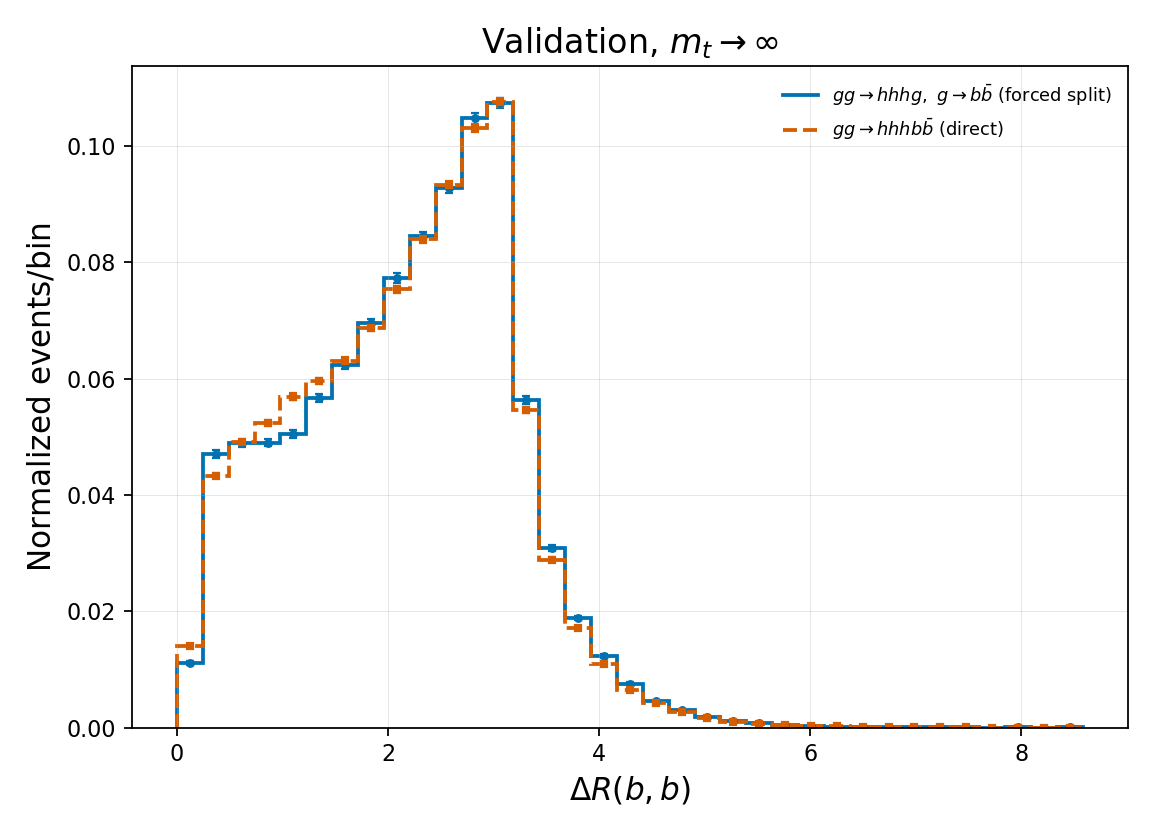}
  \hfill
  \includegraphics[height=0.22\textheight,keepaspectratio]{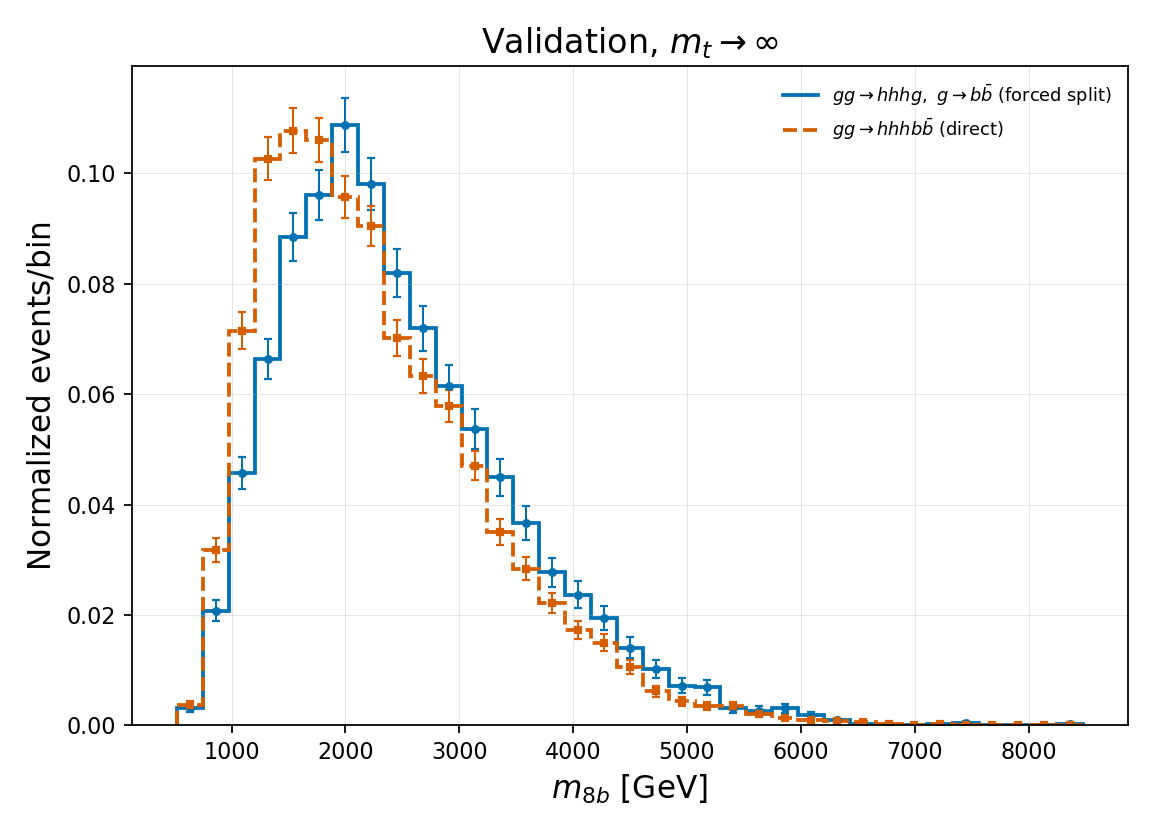}
  \caption{
    Shape comparison for the forced-splitting approximation in the large-$m_t$ limit.
    Left: pairwise angular separations $\Delta R(b,b)$ for all final-state
    $b$-quark pairs. Right: invariant mass of the final-state $8b$ system,
    $m_{8b}$. The forced-splitting sample is generated as $gg\to hhhg$,
    followed by $g\to b\bar b$ in the \texttt{HERWIG} parton shower, while the
    direct reference sample is generated as $gg\to hhh b\bar b$ in
    \texttt{MG5\_aMC}. Each distribution is independently normalized to unit
    area.
  }
  \label{fig:hhhbb-validation}
\end{figure*}
As a validation exercise, we compare the forced-splitting approximation to a
direct matrix-element calculation in the large-$m_t$ limit,
$m_t\to\infty$, where both $gg\to hhhg$ and $gg\to hhh b\bar b$ can be
generated in \texttt{MG5\_aMC}. The two samples are produced with the same
generation-level cuts, and the $gg\to hhhg$ sample is subsequently passed
through the \texttt{HERWIG} parton shower with a forced final-state
$g\to b\bar b$ splitting. Each displayed distribution is independently
normalized to unit area. The comparison therefore provides a qualitative
check of the kinematic shapes only and does not validate the absolute
normalization. We show the pairwise angular distances between all final-state
$b$-quarks and the invariant mass of the $8b$ system in
Fig.~\ref{fig:hhhbb-validation}. The transverse-momentum distributions are
shown in Fig.~\ref{fig:hhhbb-validation2}.
\begin{figure*}[b]
  \centering
\includegraphics[height=0.32\textheight,keepaspectratio]{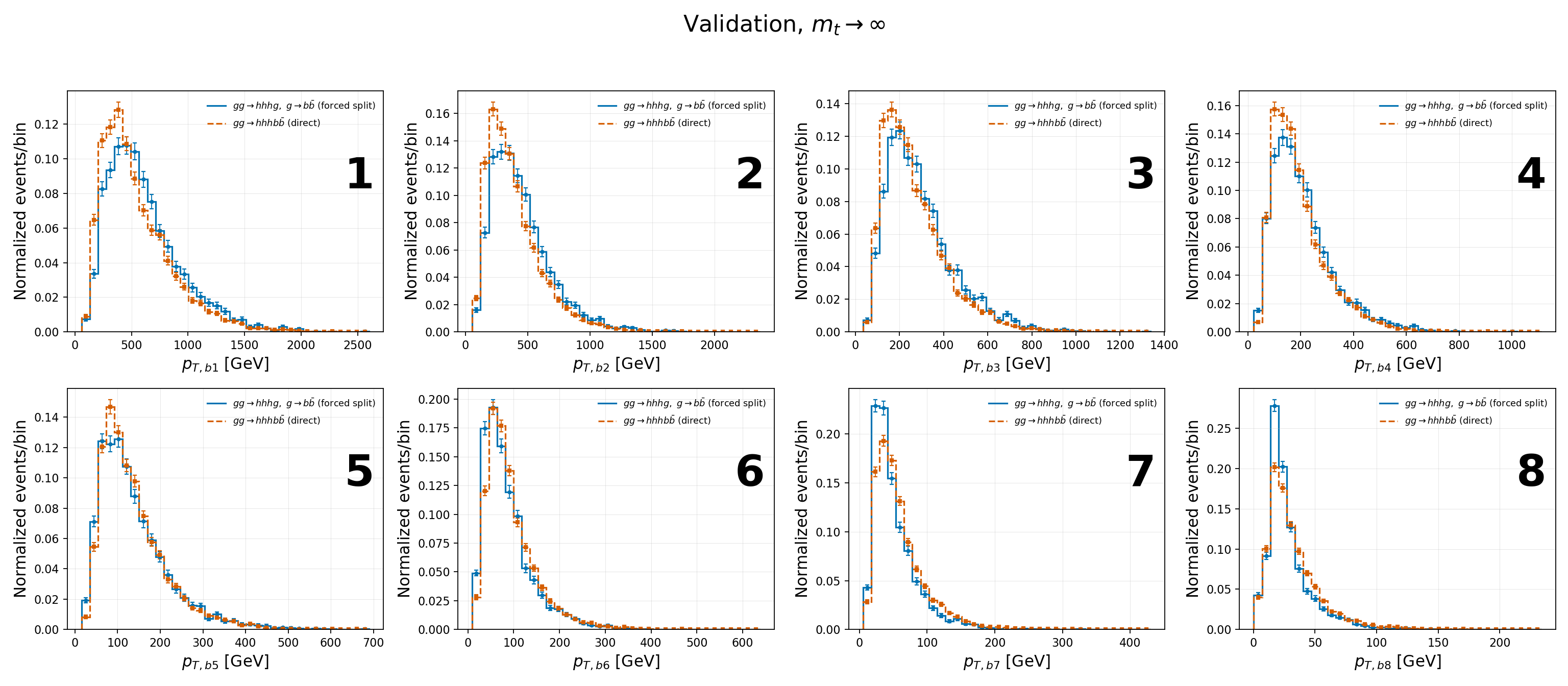}
  \caption{
    Shape comparison for the forced-splitting approximation in the large-$m_t$ limit. The independently unit-normalized transverse-momentum distributions of the eight $b$-quarks are shown in descending order. The forced-splitting sample is generated as $gg\to hhhg$,
    followed by $g\to b\bar b$ in the \texttt{HERWIG} parton shower, while the
    direct reference sample is generated as $gg\to hhh b\bar b$ in
    \texttt{MG5\_aMC}.
  }
  \label{fig:hhhbb-validation2}
\end{figure*}

\FloatBarrier
\clearpage

\section{Mathematical Definitions of the Classifier Observables}
\label{app:classifier-observables}

For completeness, we collect here the mathematical definitions of the
quantities supplied to the two classifiers. All four-vectors in this appendix
refer to the smeared jets described in Sec.~\ref{sec:jet-smearing}. For a set
of reconstructed objects \(A\), we use
\begin{equation}
\begin{aligned}
 P_A^\mu&=\sum_{i\in A}p_i^\mu, &
 m_A&=\sqrt{P_A^\mu P_{A\mu}},\\
 p_{T,A}&=\sqrt{P_{A,x}^2+P_{A,y}^2}, &
 y_A&=\frac{1}{2}\ln\frac{P_A^0+P_{A,z}}{P_A^0-P_{A,z}}\;.
\end{aligned}
\label{eq:appendix-common-kinematics}
\end{equation}
Azimuthal differences are mapped to the interval \([0,\pi]\),
\begin{equation}
 \Delta\phi_{AB}=
 \left|\operatorname{atan2}\!\left[
 \sin(\phi_A-\phi_B),\cos(\phi_A-\phi_B)\right]\right|,
\end{equation}
and the angular distance used in the reconstruction is
\begin{equation}
 \Delta R_{AB}=
 \sqrt{(y_A-y_B)^2+\Delta\phi_{AB}^{\,2}}.
\label{eq:appendix-delta-r}
\end{equation}

\subsection{Self-Coupling Analysis}
\label{app:self-coupling-observables}

Let \(j_r\), \(r=1,\ldots,8\), denote the eight selected AK4 jets, ordered by
decreasing transverse momentum. There are
\(8!/(2^4\,4!)=105\) ways to divide them into four unordered pairs. For a
pairing \(P\), the four Higgs boson candidate momenta are
\begin{equation}
 q_a^\mu(P)=p_{j_{r_a}}^\mu+p_{j_{s_a}}^\mu,
 \qquad a=1,\ldots,4.
\label{eq:appendix-resolved-candidates}
\end{equation}
The candidates are subsequently ordered by decreasing \(p_T\). The target
masses assigned in this order are
\begin{equation}
 (T_1,T_2,T_3,T_4)=(120,115,110,105)~\mathrm{GeV}.
\label{eq:appendix-mass-targets}
\end{equation}
For every pairing we define
\begin{equation}
\begin{aligned}
 m_a(P)&=m\!\left(q_a(P)\right),\\
 \delta_a(P)&=\left|m_a(P)-T_a\right|,\\
 \chi_8(P)&=\left[\sum_{a=1}^{4}\delta_a(P)^2\right]^{1/2}.
\end{aligned}
\label{eq:appendix-chi8}
\end{equation}
The pairing \(P_0\) with the smallest \(\chi_8\) defines the four reconstructed
Higgs boson candidates \(H_a\). The second-ranked distinct pairing is denoted
by \(P_1\). The pairing-quality observables are therefore
\begin{equation}
 \chi_8^{(0)}=\chi_8(P_0),\qquad
 \chi_8^{(1)}=\chi_8(P_1),\qquad
 \Delta\chi_8=\chi_8^{(1)}-\chi_8^{(0)},
\label{eq:appendix-pairing-quality}
\end{equation}
together with
\begin{equation}
 N_{60}=\sum_{P}
 \Theta\!\left(60~\mathrm{GeV}-\chi_8(P)\right),
\label{eq:appendix-n60}
\end{equation}
where the sum runs over all 105 pairings.

Writing \(q_a=q_a(P_0)\), the candidate-level inputs are
\begin{equation}
\begin{aligned}
 m_a&=\sqrt{q_a^2},&
 p_{T,a}&=p_T(q_a),\\
 \Delta R_{bb,a}
 &=\Delta R(j_{r_a},j_{s_a}),&
 z_a&=\frac{\min(p_{T,j_{r_a}},p_{T,j_{s_a}})}
 {p_{T,j_{r_a}}+p_{T,j_{s_a}}},
\end{aligned}
\label{eq:appendix-self-candidate-inputs}
\end{equation}
for \(a=1,\ldots,4\), along with the four residuals
\(\delta_a=\delta_a(P_0)\). For each of the six pairs \(a<b\), we use
\begin{equation}
 m_{ab}=\sqrt{(q_a+q_b)^2},
 \qquad \Delta R_{ab}=\Delta R(q_a,q_b),
\label{eq:appendix-self-pair-inputs}
\end{equation}
and for each of the four triples \(a<b<c\),
\begin{equation}
 m_{abc}=\sqrt{(q_a+q_b+q_c)^2}.
\label{eq:appendix-self-triple-inputs}
\end{equation}
Finally, with \(Q^\mu=\sum_{a=1}^{4}q_a^\mu=\sum_{r=1}^{8}p_{j_r}^\mu\), the
whole-event inputs are
\begin{equation}
\begin{aligned}
 m_{8b}&=\sqrt{Q^2},&
 H_T^{8b}&=\sum_{r=1}^{8}p_{T,j_r},\\
 R_{p_T}&=\frac{p_T(Q)}{\sqrt{Q^2}},&
 Y_{4h}&=|y_Q|.
\end{aligned}
\label{eq:appendix-self-event-inputs}
\end{equation}
Thus the complete set consists of the eight ordered jet transverse momenta;
\(m_{8b}\); the four values of each of
\(\delta_a,m_a,p_{T,a},\Delta R_{bb,a}\) and \(z_a\); the four pairing-quality
quantities in Eqs.~\eqref{eq:appendix-pairing-quality}
and~\eqref{eq:appendix-n60}; the six values of each of \(m_{ab}\) and
\(\Delta R_{ab}\); the four \(m_{abc}\) values; and the remaining three
whole-event quantities \(H_T^{8b}\), \(R_{p_T}\) and \(Y_{4h}\). These are the
52 inputs summarized in Table~\ref{tab:xgboost-observables}.

\subsection{Direct and Cascade Resonance Analyses}
\label{app:resonance-observables}

The resonant classifiers use the same target masses in
Eq.~\eqref{eq:appendix-mass-targets}, but allow each Higgs boson candidate to
be either resolved or merged. For a resolved candidate constructed from two
AK4 jets and a merged candidate constructed from a tagged AK8 jet \(J\), we
define, respectively,
\begin{equation}
\begin{array}{lll}
 q_a^\mu=p_{j_r}^\mu+p_{j_s}^\mu,&
 \widetilde m_a=\sqrt{q_a^2},&t_a=0,\\[2mm]
 q_a^\mu=p_J^\mu,&
 \widetilde m_a=m_{\mathrm{SD}}(J),&t_a=1.
\end{array}
\label{eq:appendix-hybrid-candidates}
\end{equation}
Here \(q_a\) always uses the ungroomed four-vector, while
\(\widetilde m_a\) is the mass used to select the assignment. After ordering
the four candidates by decreasing \(p_T\), each complete assignment \(A\) is
ranked using the dimensionless quantity
\begin{equation}
 D_4(A)=\sum_{a=1}^{4}
 \left(\frac{\widetilde m_a(A)-T_a}{125~\mathrm{GeV}}\right)^2.
\label{eq:appendix-resonance-pairing-score}
\end{equation}
The common denominator does not change which assignment is selected. We
retain the smallest and second-smallest values,
\begin{equation}
 D_4^{(0)},\qquad D_4^{(1)},\qquad
 \Delta D_4=D_4^{(1)}-D_4^{(0)}.
\label{eq:appendix-resonance-score-gap}
\end{equation}

Let \(N_{\mathrm{AK8}}\leq4\) be the number of leading AK8 jets considered in
the analytic pass/fail construction and \(n_{\mathrm{m}}\) the number that
pass the double-\(b\) tag in a given tagging outcome. The reconstruction
category is encoded as
\begin{equation}
 c(n_{\mathrm{m}})=
 \begin{cases}
 0,&n_{\mathrm{m}}=0\quad\text{(resolved)},\\
 1,&n_{\mathrm{m}}=1,2\quad\text{(mixed)},\\
 2,&n_{\mathrm{m}}=3,4\quad\text{(boosted)}.
 \end{cases}
\label{eq:appendix-resonance-category}
\end{equation}
For each tagging outcome, the common set of reconstructed inputs is
\begin{equation}
\begin{aligned}
 \mathcal O_{\mathrm{base}}=\{&
 N_{\mathrm{AK8}},n_{\mathrm{m}},c,
 D_4^{(0)},D_4^{(1)},\Delta D_4;\\
 &p_T(j_k),\quad k=1,\ldots,8;\\
 &m_a,\widetilde m_a,p_{T,a},y_a,t_a,
 \quad a=1,\ldots,4;\\
 &p_T(J_r),\eta(J_r),m(J_r),\quad r=1,\ldots,4;\\
 &m_{\mathrm{SD}}(J_r),\tau_{21}(J_r),
 \quad r=1,\ldots,4;\\
 &m_{ab},\Delta R_{ab},|\Delta y_{ab}|,\Delta\phi_{ab},
 \quad a<b;\\
 &m_{4h},p_{T,4h},y_{4h},H_T,\mathcal C,S_T\}.
\end{aligned}
\label{eq:appendix-resonance-base-set}
\end{equation}
In this expression,
\begin{equation}
\begin{aligned}
 m_a&=\sqrt{q_a^2},&
 m_{ab}&=\sqrt{(q_a+q_b)^2},\\
 \Delta y_{ab}&=y_a-y_b,&
 \tau_{21}&=\frac{\tau_2}{\tau_1},
\end{aligned}
\label{eq:appendix-resonance-candidate-kinematics}
\end{equation}
and \(m_{4h}\), \(p_{T,4h}\) and \(y_{4h}\) are obtained from
\(Q^\mu=\sum_aq_a^\mu\). The scalar transverse momentum sum and centrality are
\begin{equation}
 H_T=\sum_{o}p_{T,o},\qquad
 \mathcal C=\frac{\sum_op_{T,o}}{\sum_oE_o},
\label{eq:appendix-centrality}
\end{equation}
where \(o\) runs over the resolved AK4 jets and tagged AK8 jets used in the
reconstruction. To define the transverse sphericity, we form
\begin{equation}
 \mathsf M_T=\sum_o
 \begin{pmatrix}
 p_{x,o}^2&p_{x,o}p_{y,o}\\
 p_{x,o}p_{y,o}&p_{y,o}^2
 \end{pmatrix}.
\label{eq:appendix-sphericity-tensor}
\end{equation}
If \(\lambda_1\geq\lambda_2\) are its eigenvalues, then
\begin{equation}
 S_T=\frac{2\lambda_2}{\lambda_1+\lambda_2}.
\label{eq:appendix-transverse-sphericity}
\end{equation}
The ordered AK4 and AK8 arrays are padded to eight and four entries,
respectively; unavailable transverse-momentum and mass entries are treated as
missing values by the classifier. The set in
Eq.~\eqref{eq:appendix-resonance-base-set} contains 84 reconstructed inputs.

The direct classifier receives eight additional quantities for each tested
mass \(M_S\):
\begin{equation}
 \mathcal O_{\mathrm{dir}}^{\mathrm{mass}}=
 \left\{
 \frac{M_S}{5000~\mathrm{GeV}},
 \frac{m_{4h}-M_S}{M_S},
 \frac{m_{ab}}{M_S}\ \text{for all }a<b
 \right\}.
\label{eq:appendix-direct-mass-inputs}
\end{equation}
It therefore uses 92 inputs in total.

For the cascade classifier, the four Higgs boson candidates can be grouped
into two intermediate-scalar candidates in three ways,
\begin{equation}
 \Pi_0=(12,34),\qquad
 \Pi_1=(13,24),\qquad
 \Pi_2=(14,23).
\label{eq:appendix-cascade-pairings}
\end{equation}
For \(\Pi_r=(ab,cd)\), we define
\begin{equation}
 R_r=
 \left(\frac{m_{ab}-M_2}{M_2}\right)^2+
 \left(\frac{m_{cd}-M_2}{M_2}\right)^2.
\label{eq:appendix-cascade-score}
\end{equation}
Let \(r_\star\) denote the pairing with the smallest value, let
\(R^{(0)}\leq R^{(1)}\) be the smallest and second-smallest values, and let
\((\mu_1,\mu_2)=(m_{ab},m_{cd})\) for \(\Pi_{r_\star}\). The 14
cascade-specific inputs are
\begin{equation}
\begin{aligned}
\mathcal O_{\mathrm{cas}}^{\mathrm{mass}}=\{&
 \frac{M_2}{2400~\mathrm{GeV}},
 \frac{M_3}{5000~\mathrm{GeV}},
 \frac{M_2}{M_3};\\
 &\frac{M_3-2M_2}{M_3},
 \frac{m_{4h}-M_3}{M_3};\\
 &R^{(0)},R^{(1)},R^{(1)}-R^{(0)};\\
 &\frac{\mu_1-M_2}{M_2},
 \frac{\mu_2-M_2}{M_2},
 \frac{|\mu_1-\mu_2|}{M_2};\\
 &\frac{\mu_1}{M_2},
 \frac{\mu_2}{M_2},
 r_\star\}.
\end{aligned}
\label{eq:appendix-cascade-mass-inputs}
\end{equation}
The cascade classifier therefore uses 98 inputs in total. The mass-hypothesis
quantities in Eqs.~\eqref{eq:appendix-direct-mass-inputs}
and~\eqref{eq:appendix-cascade-mass-inputs} allow the two classifiers to follow
the changing resonance kinematics across their scans. Event identifiers,
event weights, truth-flavour information and the probabilities assigned to
the analytic tagging outcomes are not supplied as classifier inputs.

\clearpage

\section{Classifier-Score Diagnostics for the Self-Coupling Analysis}
\label{app:self-coupling-score-diagnostics}

This appendix shows two checks of the five-bin construction described in
Sec.~\ref{sec:self-coupling-likelihood} for the self-coupling analysis. These figures support the analysis
but do not introduce any additional selection or observable.

\begin{figure*}[p]
  \centering
  \includegraphics[height=0.70\textheight,keepaspectratio]
  {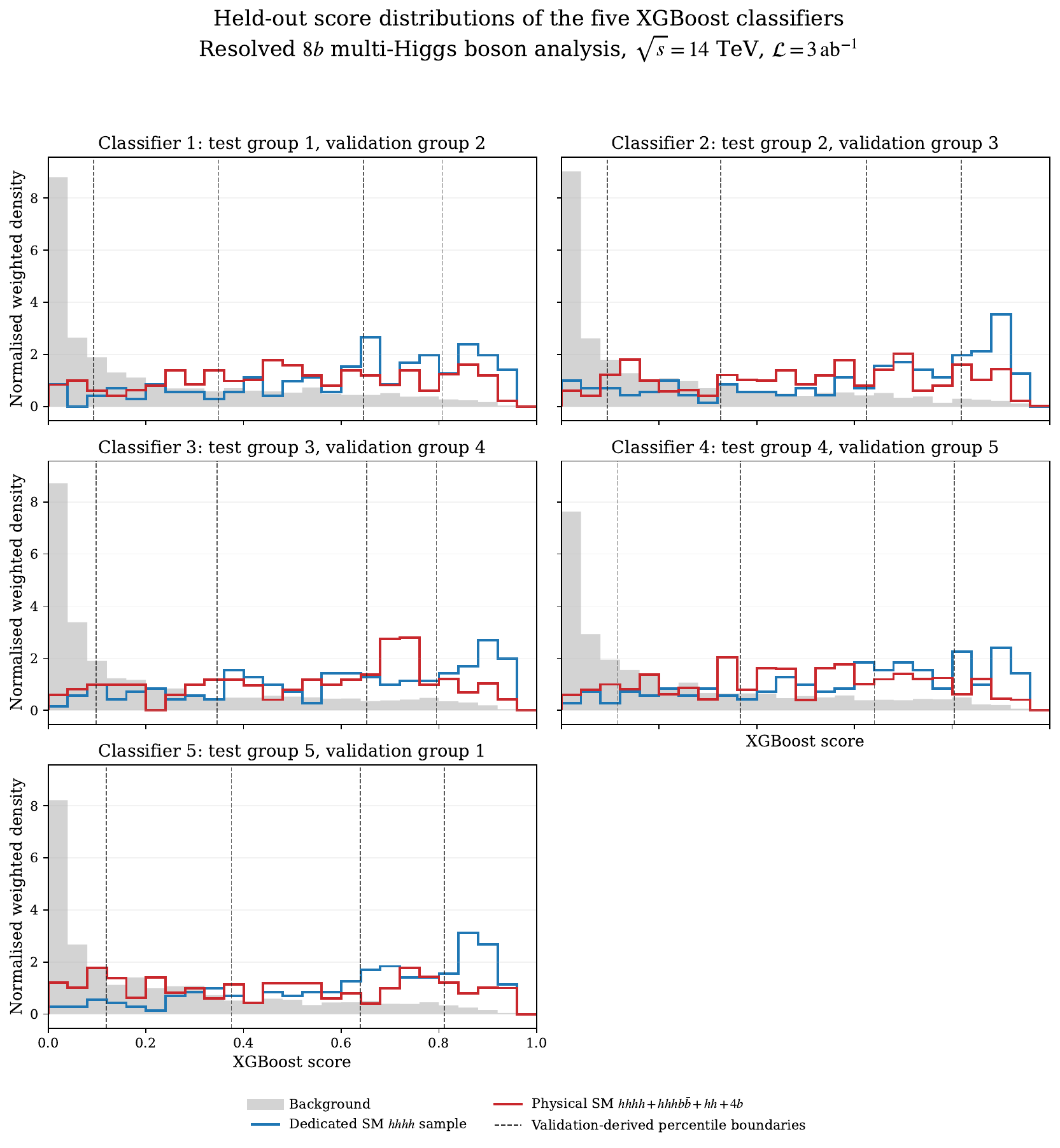}
  \caption{Classifier-score distributions obtained from the group of events
  kept aside for each of the five classifiers. The grey histogram shows the
  background, the blue curve shows the dedicated SM $hhhh$ classifier sample,
  and the red curve shows the physical SM sum of the $hhhh$, $hhh+b\bar b$, and
  $hh+4b$ processes. Each distribution is separately normalized to unit area.
  The dashed lines show the five numerical score boundaries obtained from the
  validation group corresponding to the background-percentile ranges
  \([0,0.50)\), \([0.50,0.75)\), \([0.75,0.90)\), \([0.90,0.97)\), and
  \([0.97,1]\), described in Sec.~\ref{sec:self-coupling-likelihood}.}
  \label{fig:c3d4-five-classifier-scores}
\end{figure*}

Fig.~\ref{fig:c3d4-five-classifier-scores} shows the score distributions
obtained from the events kept aside for each classifier (the held-out event
group). The grey histogram is the background, the blue curve is the dedicated
SM $hhhh$ classifier sample, and the red curve is the physical SM sum of the
$hhhh$, $hhh+b\bar b$, and $hh+4b$ processes. Each distribution is separately
normalized to unit area, so the figure compares shapes rather than predicted
event numbers. Events farther to the right have a more signal-like classifier
response. The dashed lines are the background percentile score boundaries in
Eq.~\eqref{eq:percentiles}, obtained from the corresponding validation group.
Their numerical values differ slightly because each classifier is constructed
from a different set of events, but in every panel they represent the same
five background-percentile ranges.

\begin{figure*}[p]
  \centering
  \includegraphics[width=0.91\textwidth]
  {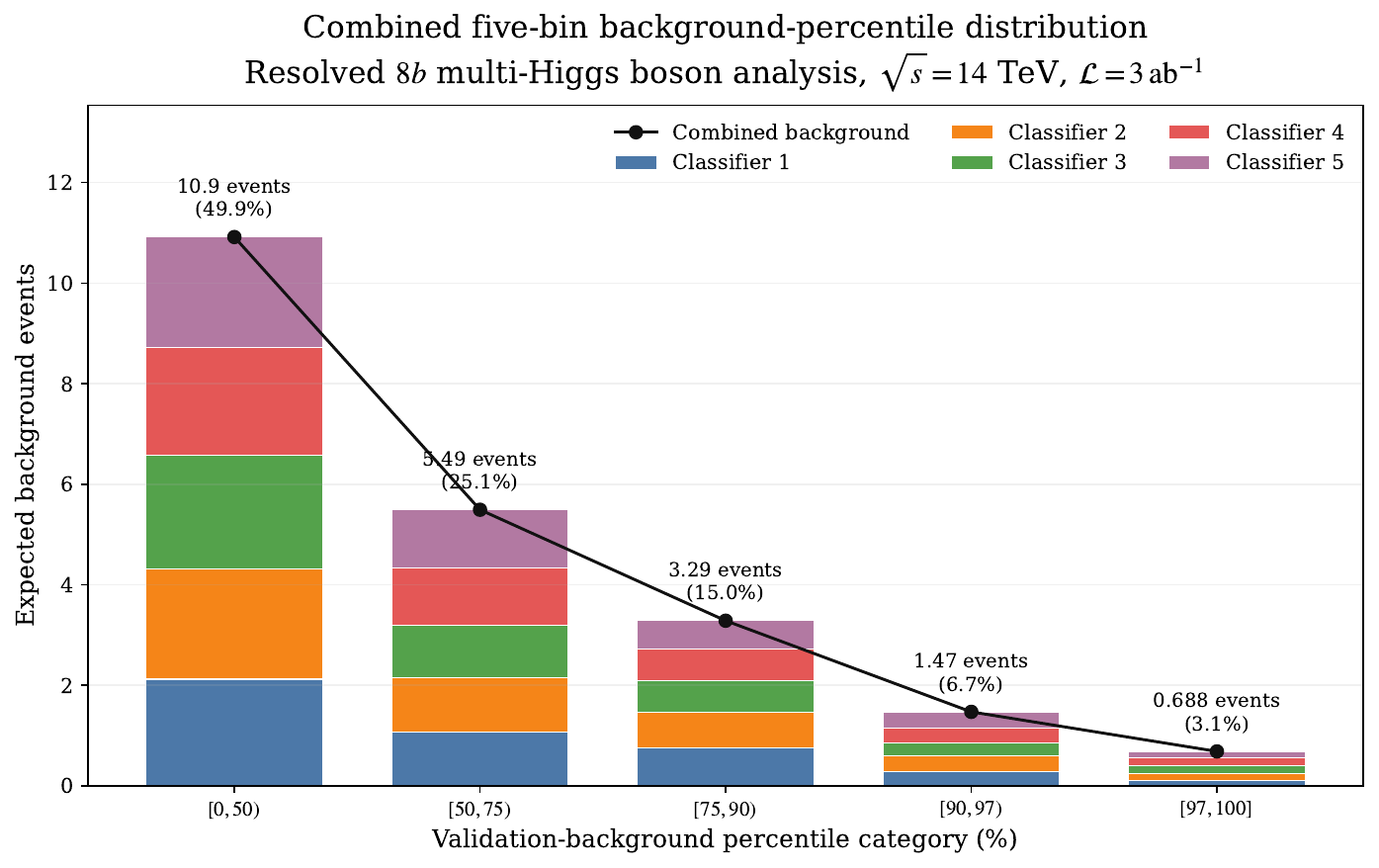}
  \caption{Combination of the five background-percentile categories. Each
  coloured part of a bar is the background contribution evaluated by one
  classifier, and the black markers show the sum over all five classifiers.
  The horizontal categories are percentiles of the background distribution in
  the validation group, not numerical classifier-score intervals. The five
  black totals are the background event counts that enter the likelihood.}
  \label{fig:c3d4-combined-percentile-background}
\end{figure*}

We next apply the five sets of boundaries to the corresponding events kept
aside and add the yields in matching percentile ranges. This produces the
single five-bin distribution shown in
Fig.~\ref{fig:c3d4-combined-percentile-background}. Each coloured part of a
bar is the contribution evaluated by a different classifier, while the black
marker is the sum over all five classifiers. From the lowest to the highest
background-percentile range, the totals are $10.92$, $5.49$, $3.29$, $1.47$,
and $0.688$ expected events. They account for $49.9\%$, $25.1\%$, $15.0\%$,
$6.7\%$, and $3.1\%$ of the full background, close to the intended fractions
of Eq.~\eqref{eq:percentiles}. The agreement is useful because the boundaries
were obtained from validation events, whereas these counts come from the
separate events kept aside. The five black totals, rather than the coloured
contributions, are the background event counts used in the likelihood. The
signal yields are combined across the five classifiers in the same way. The last bin, with 0.688 background events, corresponds to the region shown in Table~\ref{tab:self-coupling-score-yields}.

\clearpage
\bibliographystyle{JHEP}
\bibliography{biblio}

\end{document}